\documentclass[12pt, a4paper]{article}
\usepackage[scale={.75,.8}]{geometry}
\usepackage{pstricks}
\usepackage[latin2]{inputenc}
\usepackage{epic}
\usepackage{latexsym}
\usepackage{epsf}
\usepackage{epsfig}
\usepackage{amssymb,amsmath}
\usepackage{amsfonts}
\usepackage{dsfont}
\usepackage{graphicx}
\usepackage{epstopdf}
\usepackage{graphics}
\usepackage{subfigure}
\usepackage{multirow}
\usepackage{amssymb} 
\usepackage{ulem}
\usepackage{amsmath}
\usepackage{bbm} 
\usepackage{mathrsfs}   
\usepackage{xspace}

\usepackage{cite}

\newcommand{\beq}{\begin{equation}} 
\newcommand{\eeq}{\end{equation}} 
\newcommand{\ba}{\begin{array}}  
\newcommand{\ea}{\end{array}} 
\newcommand{\bea}{\begin{eqnarray}}  
\newcommand{\eea}{\end{eqnarray} }  
\newcommand{\be}{\begin{eqnarray}}  
\newcommand{\ee}{\end{eqnarray} }  
\newcommand{\bal}{\begin{align}}
\newcommand{\eal}{\end{align}}   
\newcommand{\bi}{\begin{itemize}}  
\newcommand{\ei}{\end{itemize}}  
\newcommand{\ben}{\begin{enumerate}}  
\newcommand{\een}{\end{enumerate}}  
\newcommand{\bc}{\begin{center}}
\newcommand{\ec}{\end{center}} 
\newcommand{\bt}{\begin{table}}
\newcommand{\et}{\end{table}}  
\newcommand{\btb}{\begin{tabular}}
\newcommand{\etb}{\end{tabular}}

\newcommand{\nn}{\nonumber}

\newcommand{\hu}{\ensuremath{H_{u}}}
\newcommand{\hd}{\ensuremath{H_{d}}}

\newcommand{\singlet}{\ensuremath{S}}

\newcommand{\eVdist}{\kern-0.06em}

\newcommand{\Gev}{\text{Ge\eVdist V}}
\newcommand{\Tev}{\text{Te\eVdist V}}

\newcommand{\gev}{\:\text{Ge\eVdist V}}
\newcommand{\tev}{\:\text{Te\eVdist V}}

\newcommand{\SARAH}{{\tt SARAH}\xspace}
\newcommand{\SPheno}{{\tt SPheno}\xspace}
\newcommand{\MO}{{\tt MicrOmegas}\xspace}

\newcommand{\Z}[1]{\ensuremath{\mathbbm{Z}_{#1}}} 

\begin{document}

\begin{titlepage}

\vspace*{-3.0cm}
\begin{flushright}
OUTP-13-17P\\
CERN-PH-TH/2013-186\\
DESY-13-140
\end{flushright}

\begin{center}
{\Large\bf
  Non-universal gaugino masses and fine tuning implications for SUSY searches in the MSSM  and the  GNMSSM
}

\vspace{1cm}

\textbf{
Anna Kaminska$^{a,b}$,
Graham G.~Ross$^a$,
Kai Schmidt-Hoberg$^c$
}
\\[5mm]
\textit{$^a$\small
Rudolf Peierls Centre for Theoretical Physics, University of Oxford,\\
1 Keble Road, Oxford OX1 3NP, UK
}
\\[5mm]
\textit{$^b$\small
DESY, Notkestrasse 85, D-22607, Hamburg, Germany
}
\\[5mm]
\textit{$^c$\small
Theory Division, CERN, 1211 Geneva 23, Switzerland
}
\end{center}

\vspace{1cm}

\begin{abstract}
For the case of the MSSM and the most general form of the NMSSM (GNMSSM) we determine the reduction in the fine tuning that follows from allowing gaugino masses to be non-degenerate at the unification scale, taking account of the LHC8 bounds on SUSY masses, the Higgs mass bound, gauge coupling unification and the requirement of an acceptable dark matter density. We show that low-fine tuned points fall in the region of gaugino mass ratios predicted by specific unified and string models. For the case of the MSSM the minimum fine tuning is still large, approximately 1:60 allowing for a 3 GeV uncertainty in the Higgs mass (1:500 for the central value), but for the GNMSSM it is below 1:20. We find that the spectrum of SUSY states corresponding to the low-fine tuned points in the GNMSSM is often compressed, weakening the LHC bounds on coloured states. The prospect for testing the remaining low-fine-tuned regions at LHC14 is discussed.
\end{abstract}

\end{titlepage}

\section{Introduction}\label{s.Intro}

Low energy supersymmetry 
(SUSY) was introduced as a way to solve the hierarchy problem, allowing for a consistent separation of the electroweak scale from high scales such as the Grand Unified scale, the string scale and the Planck scale. As a bonus the simplest supersymmetric extension of the Standard Model (SM), the Minimal Supersymmetric Standard Model (MSSM) \cite{Dimopoulos:1981zb}, predicts unification of the gauge couplings (to within a few percent) at a scale $M_X\sim 10^{16} \gev$  \cite{Ibanez:1981yh,Dimopoulos:1981yj}
and, together with R-parity conservation, provides a good dark matter candidate in the form of a neutralino. In addition radiative breaking naturally explains why the electroweak group and not QCD is spontaneously broken and offers an explanation for the observed scale of electroweak symmetry breaking (EWSB), relating it to the SUSY breaking scale \cite{Ibanez:1982fr,Inoue:1982pi,Inoue:1983pp,AlvarezGaume:1983gj,Ellis:1983bp}. However, despite extensive searches performed at particle accelerators and dark matter detection experiments, no direct evidence for supersymmetry has yet been observed with strong lower bounds on the masses of the coloured SUSY partners of the SM states.

In the context of models that unify the gauge symmetries\footnote{Models based on low-energy SUSY that do not address their UV completion have lower fine tuning\cite{Baer:2012cf} due to the absence of large logarithmic corrections in the RG running from the UV scale. As we are concerned with models that have gauge coupling unification, we do not consider such low-scale models here. }   these bounds already rule out much of the parameter space that has no hierarchy problem, thus losing much of the motivation for low-scale SUSY. The discovery of a Higgs-like state at approximately $125\gev$ makes the problem worse in the MSSM because the Higgs mass can only be driven this high through large radiative corrections that in turn require a large average stop mass and/or a large off diagonal entry of the stop mass matrix  $X_t=A_t-\mu\cot\beta$, both of which exacerbate the hierarchy problem.

It has been pointed out that the hierarchy problem is reduced in extensions of the MSSM\cite{Cassel:2009ps}, such as the NMSSM\cite{Dermisek:2007yt} with an additional singlet super field and particularly in its generalised version (GNMSSM)\cite{Ross:2011xv}. This has been explored in detail for the simplified case of universal boundary conditions for the SUSY breaking parameters  (CGNMSSM)\cite{Ross:2012nr}. However, even allowing for the additional contribution to the Higgs mass coming from the singlet couplings, the parameter space of this model corresponding to low fine tuning has essentially been ruled out by a combination of the non-observation at the LHC of SUSY and dark matter (DM) abundance. In particular the DM abundance has to be reduced below the ``over-closure'' limit and this is dominantly through stau co-annihilation that is only effective for relatively low $m_{0}$ and $m_{1/2}$ and hence for sparticle masses in the reach of LHC8. For more general initial conditions, in particular non-universal gaugino masses, the situation changes because the LSP can now have significant Wino/higgsino components to ensure its efficient annihilation. 

Given the need for large fine tuning in the simplest SUSY schemes, it is perhaps timely to investigate non-universal boundary conditions for gaugino masses. It has been observed that this can reduce the hierarchy problem through the appearance of a new ``focus point''  that makes the Higgs mass less sensitive to the gaugino mass scale\cite{Horton:2009ed,Choi:2005hd,Choi:2006xb,Lebedev:2005ge,Badziak:2012yg}.  Here we explore this possibility in detail for both the MSSM and GNMSSM in the context of the latest LHC bounds on SUSY states and the measurement of the Higgs mass.

Of course any considerations based on the hierarchy problem must address the question as to how it should be quantified. In this context there has been recent progress \cite{Ghilencea:2012qk,Ghilencea:2013hpa,Ghilencea:2013fka,Cabrera:2008tj} showing that the ``conventional'' fine tuning measure of the hierarchy problem follows from a normal likelihood analysis in which the electroweak breaking scale is treated as an observable. Thus minimising the fine tuning measure is elevated from an aesthetic principle to an essential part of a fit to data and a probabilistic interpretation can be given to the magnitude of the fine tuning.

The constraints of Higgs mass, dark matter abundance and its non-detection together with the gaugino focus point restriction (for reduced fine tuning) and the LHC SUSY bounds largely determines the SUSY spectrum and the resulting phenomenology of the GNMSSM with non-universal gaugino masses. In particular the focus point prefers gauginos with similar masses at low energies and, as for the CGNMSSM, rather heavy singlet states. For the case that the squarks are heavier one finds a compressed spectrum for the states accessible at LHC8 that can significantly  weaken the bounds on them. As a result there remains a significant low-fine tuned region to be probed at LHC14.

The paper is organised as follows. In the next section we define the fine tuning measure used in this analysis. In Section~\ref{sec:cuts} we discuss the experimental bounds that we use for our numerical analysis. In Section~\ref{sec:mssm} we analyse the fine tuning for the case of the MSSM with non-universal gaugino masses both analytically and numerically. In the latter case we consider the limits and resulting phenomenology coming from LHC8 on SUSY states, the Higgs mass range and dark matter. In Section~\ref{sec:gnmssm}, after motivating and discussing the structure of the GNMSSM, we perform a similar analysis of its fine tuning. We present a discussion of the phenomenology and discovery potential of the GNMSSM, paying particular regard to the possibility that its spectrum is compressed. A summary and our conclusions are presented in Section~\ref{sec:conclusions}.

\section{The fine tuning measure}
\label{sec:FT}

The fine tuning measure, $\Delta_{p}$, with respect to a given independent parameter, $p$, was introduced in \cite{Ellis:1986yg, Barbieri:1987fn}, with the form
\begin{equation} 
\Delta _{p}\equiv \frac{\partial \ln
  v^{2}}{\partial \ln p} = \frac{p}{v^2}\frac{\partial v^2}{\partial p} \;,
\end{equation}
where $v$ is the electroweak scale\footnote{$v^{2}=v_{u}^{2}+v_{d}^{2}$ where $v_{u,d}$ are the up and down sector Higgs vacuum expectation values. Here we work in conventions
in which $v \simeq 246 \gev$.}.

In  \cite{Ghilencea:2012qk,Ghilencea:2013hpa,Ghilencea:2013fka} it was shown that this measure naturally appears in a likelihood fit in which the electroweak breaking scale, $v$, is treated as an observable. In this case the likelihood is suppressed by the overall fine tuning measure $\Delta_{q}$ given by 
\begin{equation}
\Delta_{q}=\big(\sum_{p}\Delta_{p}^{2}\big)^{1/2}.
\label{ftm}
\end{equation}
Thus we see that reducing the fine tuning measure rather than being an aesthetic requirement is an essential feature of a likelihood fit of SUSY to data. 

Usually one term dominates in Eq.~(\ref{ftm}) and a more commonly used fine tuning measure is given by 
\begin{equation} 
\Delta \equiv \max {\text{Abs}}\big[\Delta _{p}\big] \; .
\end{equation}
To allow for comparison with previous work we will use $\Delta$ when computing the overall fine tuning.

The quantity $\Delta^{-1}$ gives a measure of the accuracy to which independent parameters, $p$, must be tuned to get the correct electroweak breaking scale. In the analysis presented here the parameters correspond to the independent parameters in the superpotential and the soft scalar potential plus the top Yukawa coupling\footnote{We use the modified definition of fine tuning for the top-Yukawa coupling, appropriate for measured parameters \cite{Ciafaloni:1996zh}.} all defined at the unification scale and chosen to be of mass dimension 2 where appropriate, e.g.~$\mu^{2}$. From the connection of the fine tuning measure to the likelihood one finds that an acceptable likelihood requires $\Delta$ should be much less than $100$\cite{Ghilencea:2012qk,Ghilencea:2013hpa,Ghilencea:2013fka}.

\section{SUSY, Higgs and DM Cuts}
\label{sec:cuts}

In this section we briefly describe the cuts that we impose for the numerical analyses  of the MSSM and the GNMSSM.
As discussed below, the non-universal gaugino mass case in the GNMSSM often leads to a compressed SUSY spectrum with small mass differences between gauginos and the LSP that makes SUSY discovery more difficult. To account for this in a manner consistent with the non-observation of superpartners at the LHC we implemented a cut on the gluino mass as a function of the gluino-LSP mass difference as presented in \cite{ATLAS-CONF-2013-047,CMS-PAS-SUS-13-008}. 
In Fig.~7 of \cite{ATLAS-CONF-2013-047} two bounds are shown, a weaker one for decoupled squarks and a stronger one for $m_\text{squark} \sim m_\text{gluino}$. Most parameter space points of interest to us are in the intermediate regime, but to be sure they are not excluded we will use the stronger bound.
We further require the chargino and slepton masses to be above $100\gev$.
We also require that the lightest supersymmetric particle (LSP) is a neutralino which is a good dark matter candidate and its
relic density does not exceed the $5\sigma$ PLANCK \cite{Ade:2013zuv} upper bound of $\Omega h^2 \le 0.1334$.
While an under-abundance could always be compensated by the relic density of a multitude of other particles, an overabundance would require a deviation from the standard thermal history of the Universe (or at least a sufficiently
low reheating temperature, such that the dark matter candidate never reaches thermal equilibrium). 
In addition to constraints from the relic density there are constraints from dark matter direct detection searches which limit the cross section of the lightest neutralino
to nucleons.
We require that the points are consistent with dark matter direct detection constraints, in particular with the latest constraint from XENON100 \cite{Aprile:2012nq}. As we only require dark matter not to be overabundant, the inferred cross section should be multiplied with $(\Omega h^2)^\text{th} /(\Omega h^2)^\text{obs} =(\Omega h^2)^\text{th} / 0.1199$ with  $(\Omega h^2)^\text{th}$ the predicted relic density for the given point in parameter space to account for cases with underabundant neutralinos.
Finally, for the Higgs mass we take the average of the CMS and ATLAS best fit values of $125.7 \gev$ \cite{CMS-PAS-HIG-13-005} and $125.5 \gev$ \cite{ATLAS-CONF-2013-014} respectively and allow for a $3\gev$ uncertainty, $m_h = 125.6 \pm 3\gev$.

For our numerical analyses we use \SPheno \cite{Porod:2003um,Porod:2011nf} created by \SARAH \cite{Staub:2008uz,Staub:2009bi,Staub:2010jh,Staub:2012pb}.  This version performs a complete one-loop calculation of all SUSY and Higgs masses \cite{Pierce:1996zz,Staub:2010ty} and includes the dominant two-loop corrections for the scalar Higgs masses \cite{Dedes:2003km,Dedes:2002dy,Brignole:2002bz,Brignole:2001jy}. We have extended  \SPheno  by incorporating routines to calculate the fine tuning as presented in \cite{Ross:2012nr}. The dark matter relic density as well as the direct detection bounds are calculated with \MO \cite{Belanger:2006is,Belanger:2007zz,Belanger:2010pz}.

\section{The MSSM case}
\label{sec:mssm}
We start with the determination of the fine tuning in the MSSM with non-universal gaugino masses, which we write as $M_1=a \cdot m_{1/2}, M_2=b \cdot m_{1/2}$ and $M_3=m_{1/2}$.\footnote{This scenario has recently been studied in \cite{Antusch:2012gv} based on the 2012 LHC limits. However we include it here as it is important to use exactly the same methods and cuts to calculate the fine tuning when comparing the MSSM and the GNMSSM; our estimate of the fine tuning in the MSSM case is higher than in  \cite{Antusch:2012gv}.} All other parameters we take to be (C)MSSM like, i.e.\ a universal scalar mass and all other soft terms proportional to their corresponding superpotential couplings. 
\subsection{The gaugino focus point - analytic discussion}
\label{sec:MSSManalytic}
As pointed out in \cite{Horton:2009ed}, having solved the RGEs, the soft masses in the Higgs sector at a given energy scale $Q$ can be written as a polynomial in initial parameters
\bea
m_{h_u}^2(Q) & = & z_{h_u}^{m_0}(Q) m_0^2+z_{h_u}^{m_{1/2}}(Q) m^2_{1/2}+z_{h_u}^{A_0}(Q) A_0^2+2z_{h_u}^{m_{1/2}A_{0}}(Q) m_{1/2} A_0 \nn \\
m_{h_d}^2(Q) & = & z_{h_d}^{m_0}(Q) m_0^2+z_{h_d}^{m_{1/2}}(Q) m^2_{1/2}+z_{h_d}^{A_0}(Q) A_0^2+2z_{h_d}^{m_{1/2}A_0}(Q) m_{1/2} A_0 .
\label{zH}
\eea
Using this equation the relation between the electroweak VEV,  $v$, and the independent parameters defined at the unification scale can be written as~\cite{Horton:2009ed}
\begin{equation}
\lambda^{(0)}v^2=-\frac{\tan^2\beta}{\tan^2\beta-1}\bar{m}_{h_u}^2+\frac{1}{\tan^2\beta-1}\bar{m}^2_{h_d}-|\mu|^2 ,
\label{EWbreaking}
\end{equation}
where $\lambda^{(0)}$ is the tree-level quartic Higgs coupling and 
\begin{equation}
\bar{m}^2_{h_x}=m^2_{h_x}+\frac{\partial V^{(1)}}{\partial h^2_x}
\end{equation}
with $V^{(1)}$ the one-loop Coleman-Weinberg effective potential. 

For the analytic discussion we will for simplicity concentrate on the case with small $A_{0}$ such that we can drop the terms proportional to $m_{1/2} A_0$ in Eq.~(\ref{zH}).
As we will also be interested in the low $\tan \beta$ regime both $m_{h_u}^2$ and $m_{h_d}^2$ contributions are non-negligible for the derivation of the EW scale. In this case, with
\begin{equation}
z_{h_{ud}}^{m_{1/2}}(Q) = \frac{\tan^{2}\beta}{\tan^{2}\beta - 1} z_{h_u}^{m_{1/2}}(Q)- \frac{1}{\tan^{2}\beta - 1}z_{h_d}^{m_{1/2}}(Q),
\end{equation}
the gaugino focus point is defined as the energy scale at which $z_{h_{ud}}^{m_{1/2}}$ vanishes. 
Threshold corrections to the RGE running are neglected in our analytical estimates.

\begin{figure}[h*]
\centering
\includegraphics[width=0.32\linewidth]{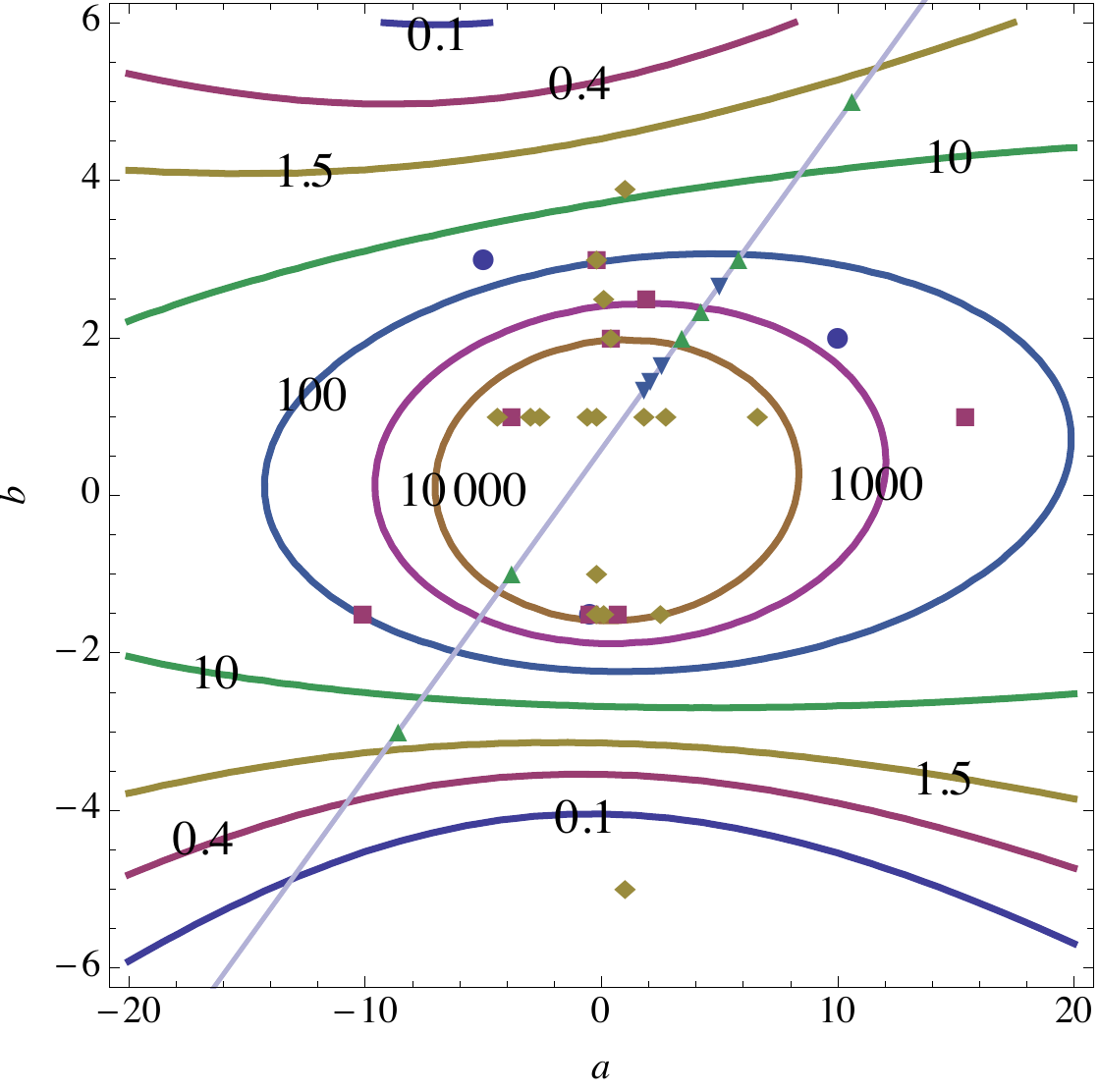}  
\includegraphics[width=0.32\linewidth]{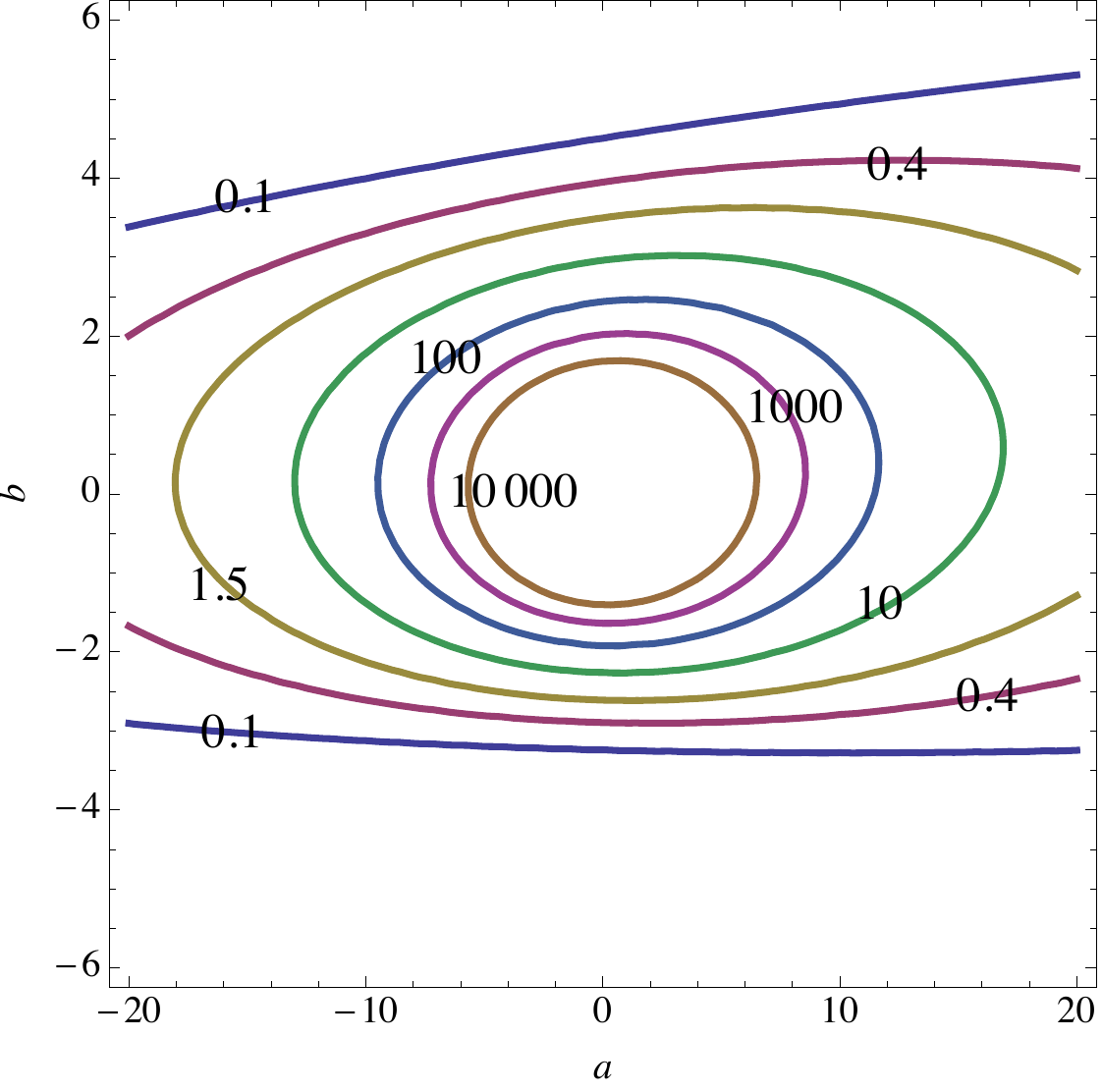}  
\includegraphics[width=0.32\linewidth]{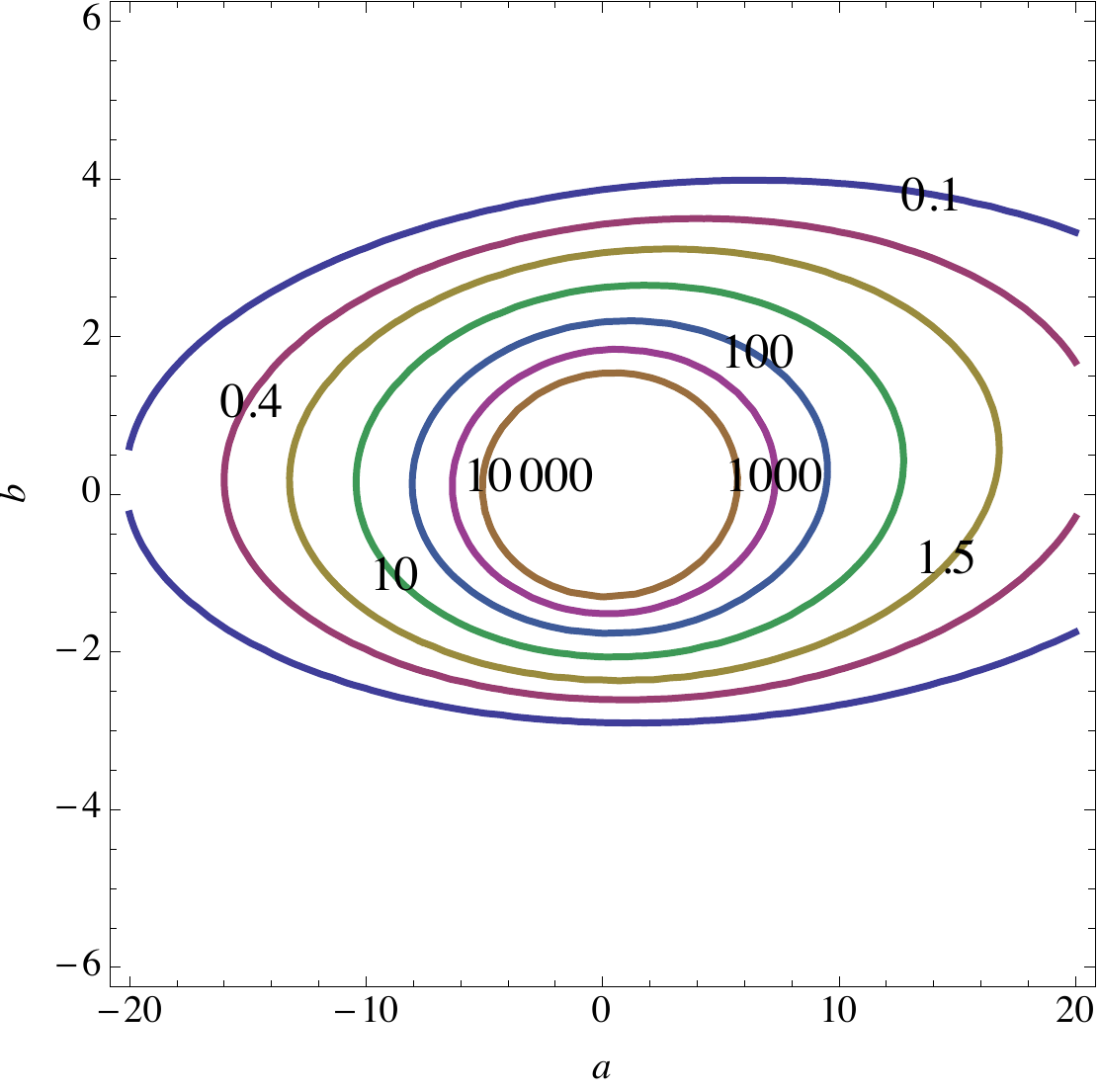}  
\caption{Analytic results for the gaugino focus point scale contours (in units of $ \Tev$) in the MSSM for $\tan \beta=2,3,10$ from left to right. 
The points and the gray line correspond to specific models with non-universal gaugino masses; see text for details. The contours do not change much for larger $\tan\beta$. }
\label{MSSManalytic}
\end{figure}

In order to compute $z_{h_{ud}}^{m_{1/2}}(Q)$ analytically we assume that only the top Yukawa is non-zero. In this case exact solutions to the RGEs in a closed form can be found following \cite{Ibanez:1983di} and \cite{Mambrini:2001wt}. The results for the gaugino focus point are given in  Fig.~\ref{MSSManalytic}. For large $\tan \beta$ they are consistent with \cite{Horton:2009ed}. For the case that the focus point is of order the electroweak breaking scale the sensitivity of $v$ to the gaugino mass scale $m_{1/2}$ is significantly reduced and so the fine tuning is reduced. 
Approximate expressions for the fine tuning in the MSSM case are given in the Appendix. 

\begin{figure}[h*]
\centering

\includegraphics[width=0.33\linewidth]{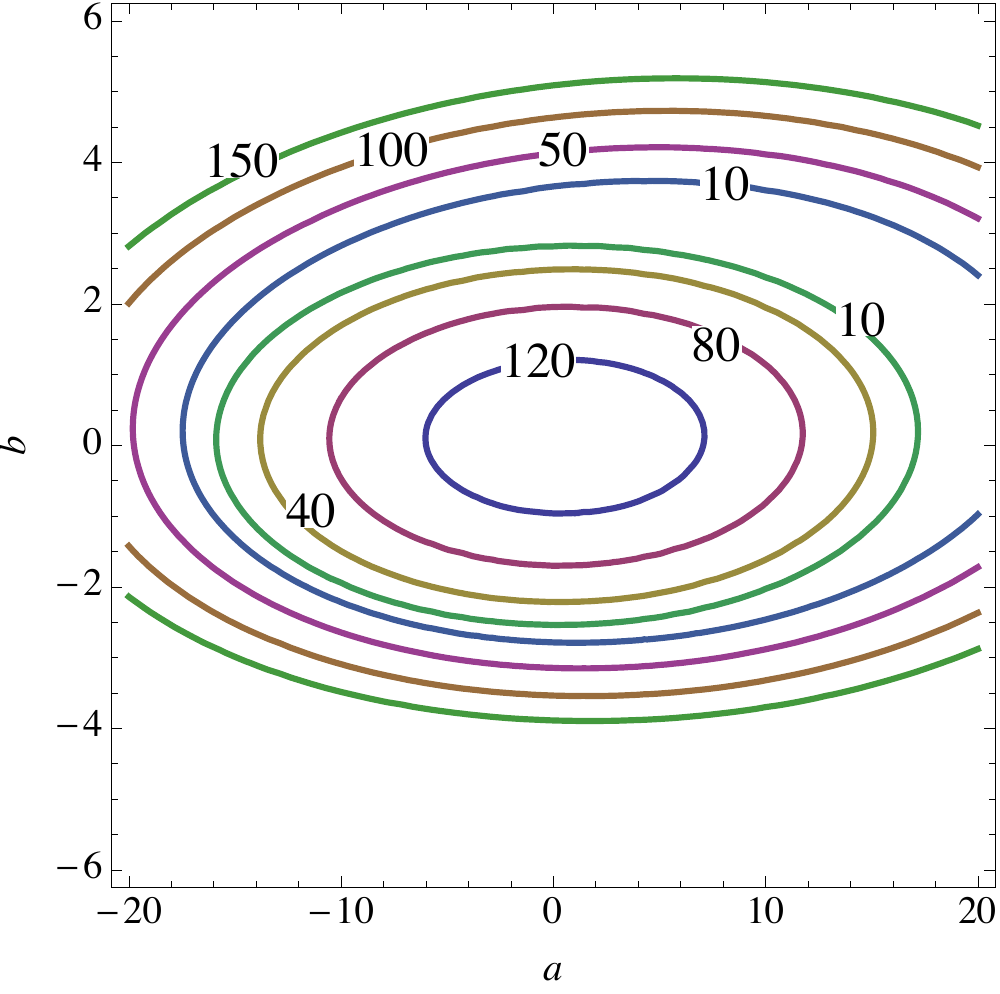}  
\hspace{1cm}
\includegraphics[width=0.33\linewidth]{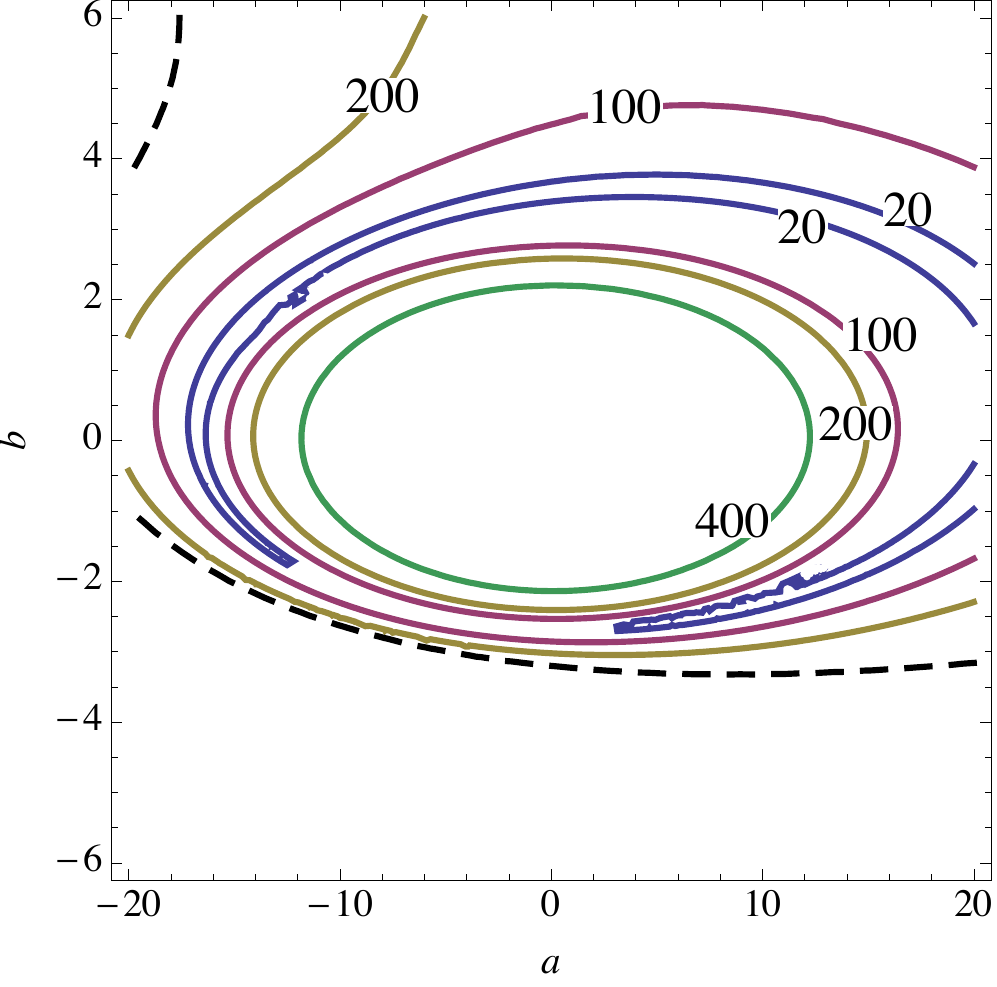}   
\caption{Analytic results for minimal fine tuning in the MSSM in the limit $\tan\beta\rightarrow\infty$ before (left) and after (right) the Higgs mass cut. The region below the black dashed line 
corresponds to very large scalar masses, $m_0 > 5 \tev$.}
\label{MSSManalyticFT1}
\end{figure}

It may be seen that the reduction of $z_{h_{ud}}^{m_{1/2}}(v)$ when the gaugino focus point approaches the electroweak scale not only reduces $\Delta_{m_{1/2}}$ but also reduces the fine tuning with respect to $\mu$. This happens because $\mu^2$ is related to $m_{h_u}^2$ and $m_{h_d}^2$ through the electroweak symmetry breaking condition Eq.~(\ref{EWbreaking}). Contours of minimal overall fine tuning for large $\tan\beta$, obtained only requiring the existence of viable electroweak breaking and $m_{1/2}>0.5\Tev$, are shown in Fig.~\ref{MSSManalyticFT1}. The situation before and after the Higgs mass cut is presented. The general behavior of fine tuning in both cases is similar but satisfying the Higgs mass cut demands larger values of $m_{1/2}$ and optimal $A_0$, which increases the fine tuning. Analytic study shows that in the central part of the considered $(a,b)$ plane $\Delta_{\mu_0}$ and $\Delta_{m_{1/2}}$ dominate the overall fine tuning. In this region the gaugino focus point is much above the electroweak scale, hence $m_{h_u}^2$ depends strongly on $m_{1/2}$ forcing large negative values of $m_{h_u}^2$ and large values of $\mu$ around the $\Tev$ scale. As the absolute values of $a$ and $b$ grow, $|z_{h_{ud}}^{m_{1/2}}(v)|$ decreases and both $\Delta_{\mu_0}$ and $\Delta_{m_{1/2}}$ become smaller. The smallest fine tuning corresponds to values of $a$ and $b$ for which the gaugino focus point scale is close to the electroweak scale. For moderate $a$, the value $|b|\sim 2.5-3$ corresponds to such a low-scale gaugino focus point. The outer region of the $(a,b)$ plane corresponds to the gaugino focus point scale below the electroweak scale. In this case $z_{h_{ud}}^{m_{1/2}}(v)$ is positive and grows as the absolute values of $a$ and $b$ increase. This means that gauginos give a positive contribution to $m_{h_u}^2$ and values of $\mu^2$ are very small. The fine tuning is dominated by $\Delta_{m_{1/2}}$ and $\Delta_{A_{0}}$. As the gaugino focus point scale moves further below the electroweak scale the value of $m_0$ needed to guarantee $m_{h_u}^2<0$ becomes large and the fine tuning increases. If we demand that $m_{0}<5\;\Tev$, values of $a$ and $b$ below the black dashed line in the right panel of Fig.~\ref{MSSManalyticFT1} are excluded.

\begin{figure}[h*]
\centering

\includegraphics[width=0.33\linewidth]{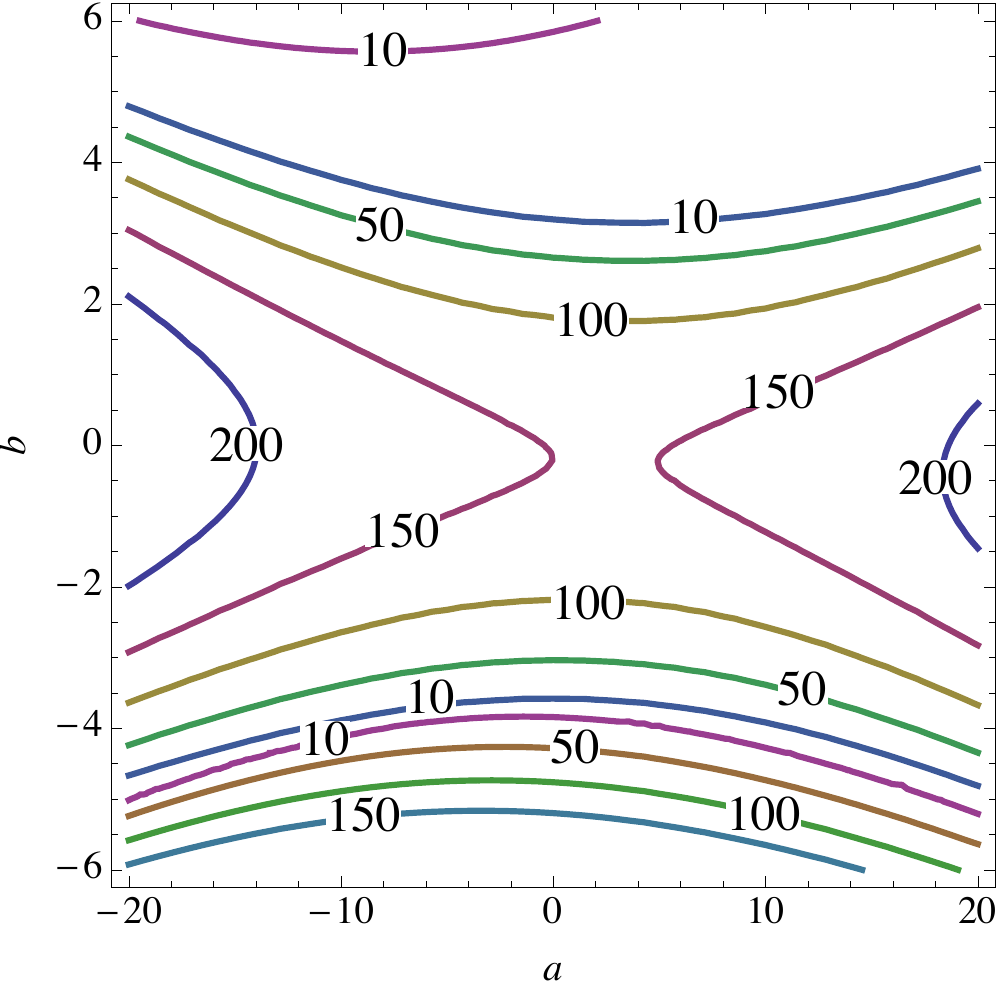} 
\hspace{1cm} 
\includegraphics[width=0.33\linewidth]{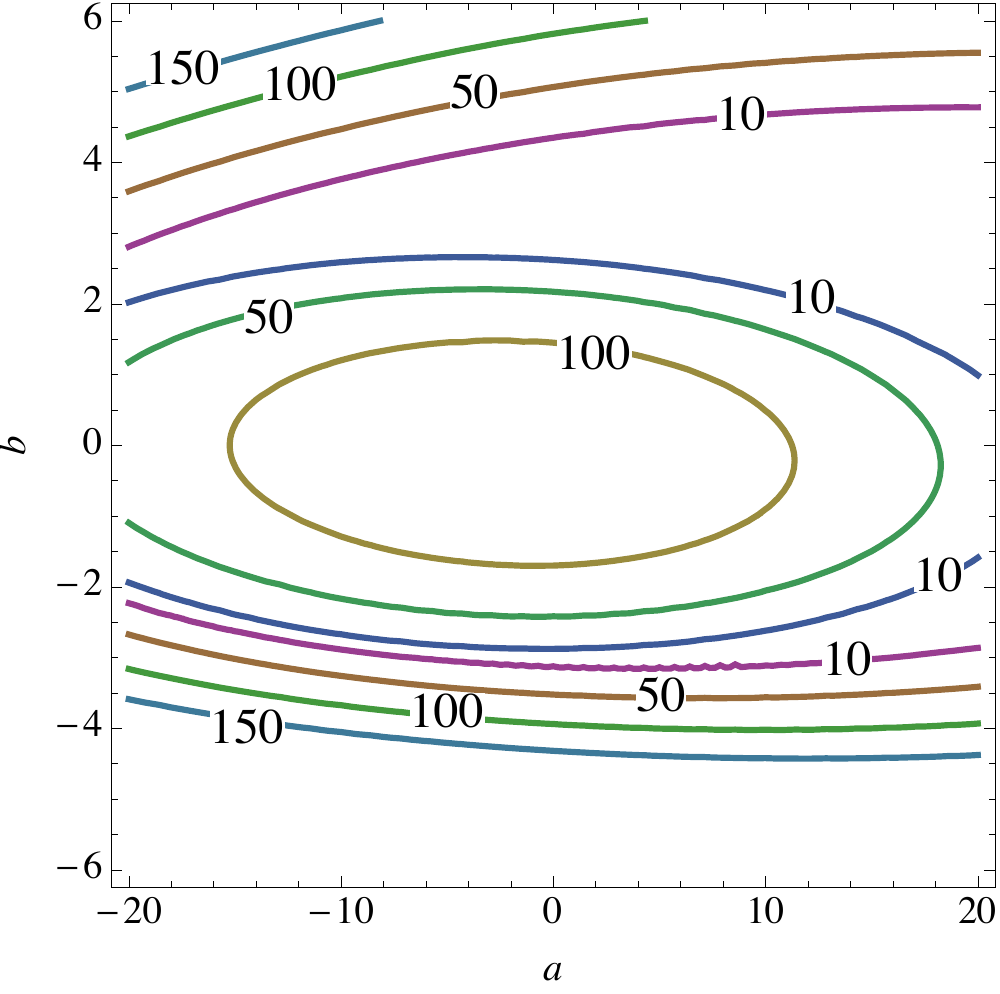}   
\caption{Analytic results for minimal fine tuning in the MSSM for $\tan \beta=2,3$ without the Higgs mass cut.}
\label{MSSManalyticFT2}
\end{figure}

Fig.~\ref{MSSManalyticFT2} shows the estimated minimal fine tuning for small values of $\tan\beta$ without imposing the Higgs mass cut (but still requiring $m_{1/2}>0.5\;\Tev$). As in the large $\tan\beta$ case, the central region of the $(a,b)$ plane corresponds to large values of $\mu^2$ and the fine tuning with respect to $\mu$ dominates, along with $\Delta_{m_{1/2}}$. The smallest fine tuning is again obtained for values of $a$ and $b$ corresponding to the gaugino focus point near the electroweak scale. The large $|b|$ region of the plot is related to small values of $\mu^2$ and the fine tuning becomes dominated by $\Delta_{m_{1/2}}$ growing as the gaugino pocus point scale moves away from the electroweak scale. The  fine tuning contours do not exactly follow the gaugino focus point scale contours because the $m_{1/2} A_0$ term in $m_{h_u}^2$ and $m_{h_d}^2$ plays a non-negligible role.

Of course, if arbitrary values of the parameters $a$ and $b$ are chosen, the contribution to the overall fine tuning from $\Delta_{a,b}$ should also be included in the analysis above, which typically spoils the improvement in fine tuning. However if $a$ and $b$ are fixed by the underlying theory such contributions are absent. As discussed in  \cite{Horton:2009ed} values of $a$ and $b$ in the low-focus-point region occur naturally in a variety of models. To illustrate this we show in the first plot of  Fig.~\ref{MSSManalytic}  the predicted points for the $SU(5),\; SO(10)$ and $E_{6}$ GUT models (denoted by circles, squares and diamonds respectively) considered in \cite{Martin:2009ad}. 
GUT models with F terms in \textbf{75} or \textbf{200} of $SU(5)$, in \textbf{210} or \textbf{770} of $SO(10)$ and in the corresponding representations of ``flipped $SO(10)$'' embedded in $E_6$ predict gaugino mass ratios in the intermediate and low fine tuning region.
Green triangles represent the OII orbifold model for various choices of the discrete Green Schwarz parameter, $\delta_{GS}$ \cite{Brignole:1993dj}. The values $\delta_{GS}=-5,-6,-7$ are optimal from the point of view of fine tuning.  
For comparison we also show points relevant for mirage mediation, where soft terms receive comparable contributions from gravity (modulus) and anomaly mediated SUSY breaking. In this case gaugino masses at the GUT scale have the following form
\begin{equation}
M_{a}=m_{3/2}\left( \varrho +b_{a}g_{a}^{2}\right) 
\end{equation}
where $g_{a}$ is the relevant gauge coupling, $b_{a}$ is its $\beta$-function coefficient, while $\varrho$ describes the relation between modulus and anomaly mediated contributions. This prescription for gaugino masses as a function of $\varrho$ generates the gray line in Fig.~\ref{MSSManalytic} in the $(a,b)$ parameter space. If $\varrho$ is  a continuous parameter there should be an additional contribution $\Delta_{\varrho}$ to the overall fine tuning. However specific string models fix the value of $\varrho$.
Four examples are shown in Fig.~\ref{MSSManalytic} by the blue inverted triangles: (i) the minimal setup of KKLT-type moduli stabilization in type II B string theory \cite{Choi:2007ka,Cho:2007fg,Choi:2005uz}, (ii) a model with vacuum uplifting via hidden sector matter superpotentials \cite{Choi:2007ka} (iii) and (iv) the Mini Landscape of orbifold compactifications in heterotic string theory \cite{Badziak:2012yg}  with $SU(4)$ and $SU(5)$ hidden sector gauge groups; the type II B string theory model with vacuum stabilisation by F-terms of hidden sector matter superpotentials predicts values of $a$ and $b$ in the low fine tuning region.

\subsection{Numerical analysis of fine tuning in the MSSM}
\begin{figure}[!h!]
\centering
\includegraphics[width=0.49\linewidth]{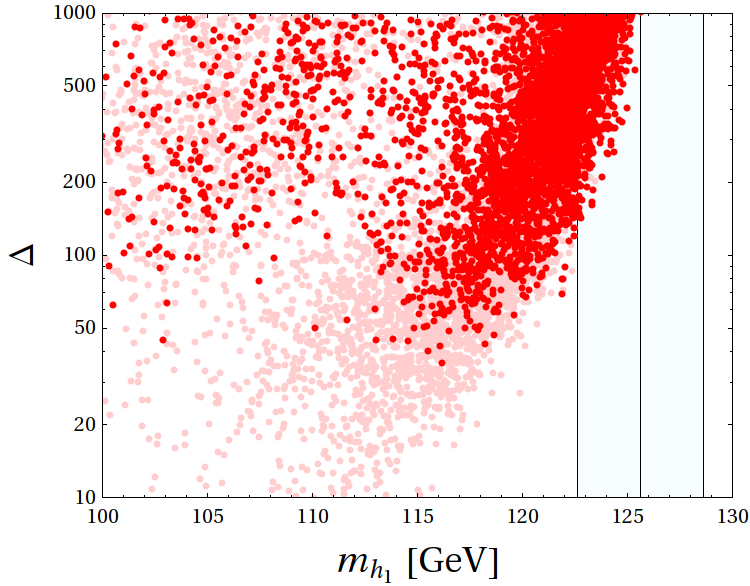} 
\includegraphics[width=0.49\linewidth]{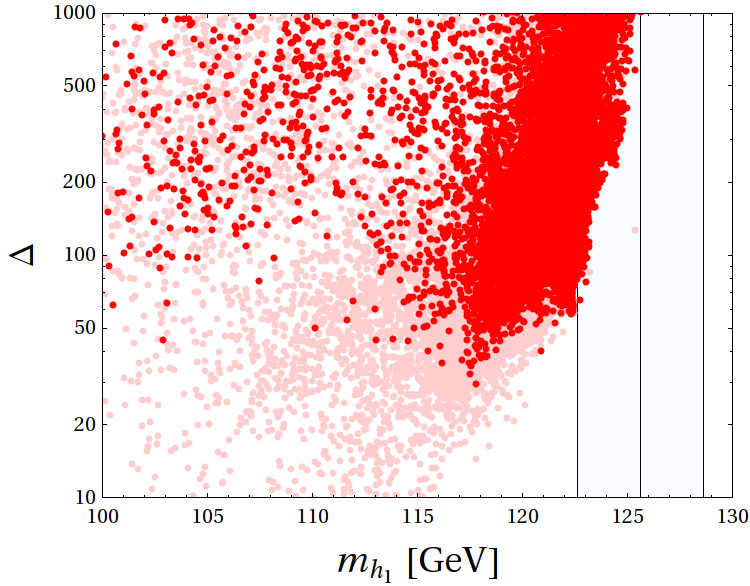} 
\caption{Dependence of the fine tuning on the Higgs mass in the MSSM.
The light red points are before any cuts while the dark red points take into account
cuts on the SUSY masses and the relic neutralino abundance. The left plot is uniform
in the density of the input parameters, their density reflects the likelihood for finding a viable point.
The right plot shows additional points where we zoomed into regions of small fine tuning. The minimal fine tuning after all cuts is about 60
for a Higgs mass of 122.6 GeV; requiring universal gaugino masses, i.e.\ $a=b=1$, it is about 350.}
\label{fig:6}
\end{figure}

We turn now to the full numerical analysis of fine tuning in the (C)MSSM for the case of non-universal gaugino masses. After imposing the electroweak symmetry breaking conditions the independent parameters, defined at the unification scale, are
$m_0$, $A_0$, $\tan \beta$, Sign $\mu$, $m_{1/2}$, $a$, $b$. 
 We compute the overall fine tuning, $\Delta$, corresponding to these parameters apart from the  contribution from $a$ and $b$ assuming, as discussed above, that they are fixed in the underlying theory. The results of a broad scan of the MSSM parameter space are shown in Fig.~\ref{fig:6}. We show the results both before and after any cuts are made as described in Section~\ref{sec:cuts}.

In order to achieve a Higgs mass in agreement with experimental results large mixing corresponding to large
A-terms is needed. Note that the fine tuning measure is large; even allowing for $\pm 3\gev$ in the Higgs mass about its central value of $125.6 \gev$ the smallest fine tuning we find after cuts on the SUSY spectrum but before the cut on the Higgs mass is about 30 but after the cut on the Higgs mass it is about 60 (if we were to take the central value without error the smallest fine tuning is about 500!).  Although still problematic, this is significantly smaller than the case of degenerate gaugino masses at the unification scale where, using the same method of evaluation, the {\it smallest} fine tuning for a $122.6\gev$ Higgs is about 350.

\begin{figure}[h*]
\centering

\includegraphics[width=0.49\linewidth]{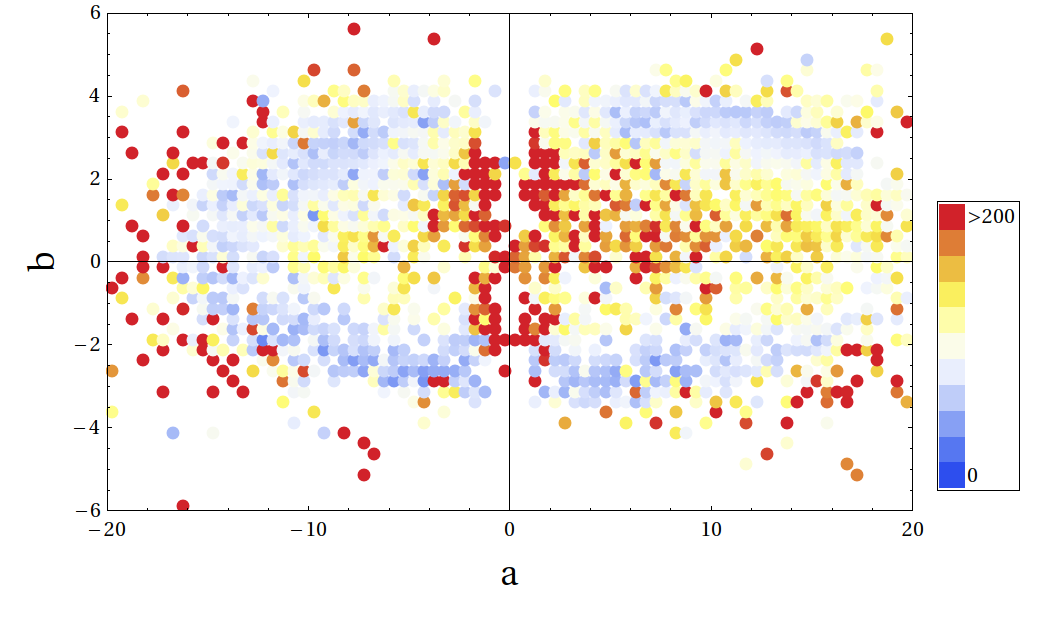}  
\includegraphics[width=0.49\linewidth]{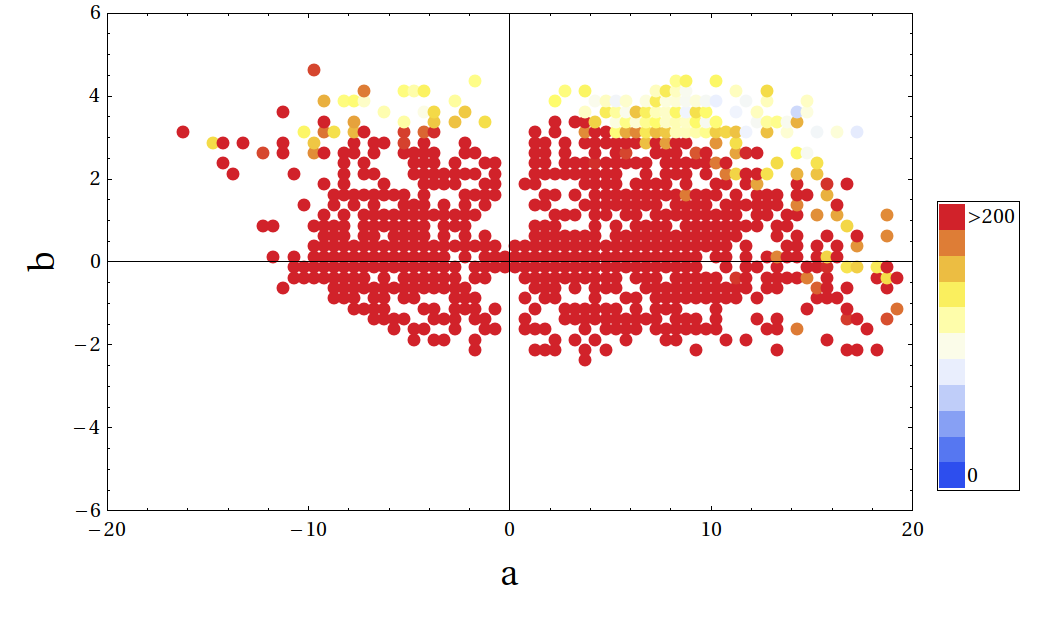}  
\includegraphics[width=0.49\linewidth]{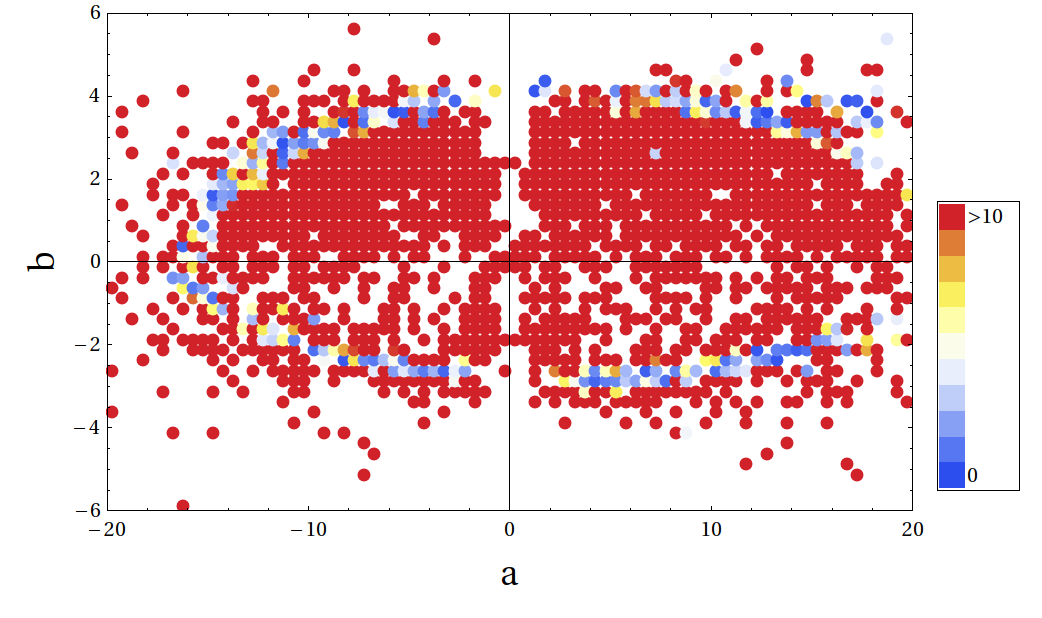}  
\includegraphics[width=0.49\linewidth]{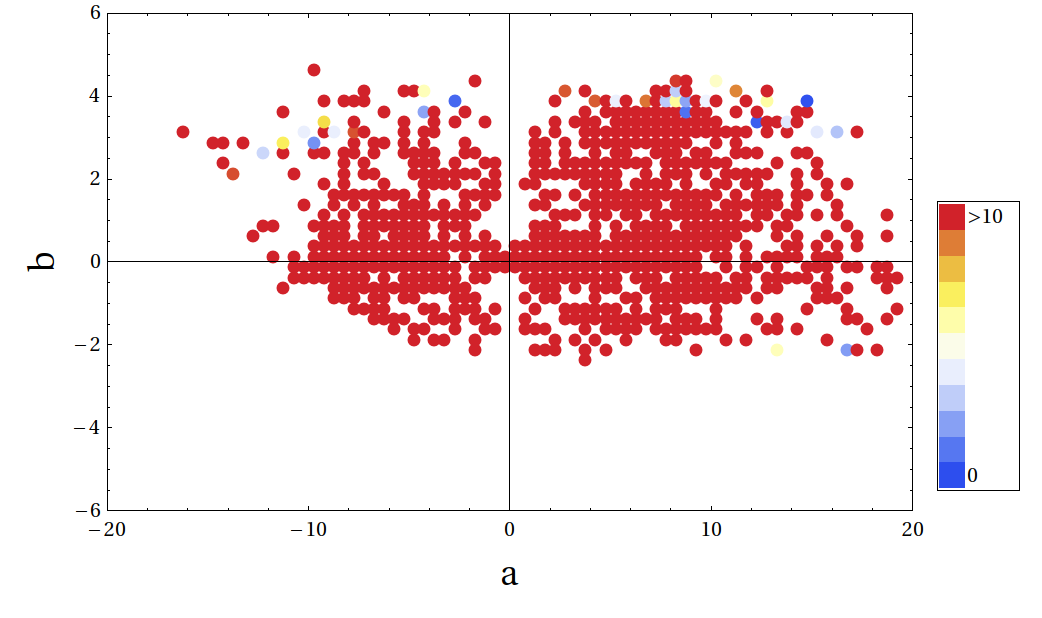}  
\caption{Dependence of the fine tuning on the input parameters $a$ and $b$ with the additional parameters chosen such that at each point 
the smallest fine tuning is realised.
The upper left plot is after SUSY and DM but before the Higgs mass cut - the upper right plot includes the cut on the Higgs mass.
It can be seen that this significantly increases the fine tuning, as expected.
To compare with the analytic estimate the plots in the lower row show the same but instead of the overall fine tuning (which is typically dominated by the fine tuning with respect to $\mu$)
the fine tuning with respect to the gaugino mass parameter $m_{1/2}$ is shown (note the change of scale). The corresponding ellipse is very close to the analytic estimate.
}
\label{fig:7}
\end{figure}

Of course it is important to check the values of the gaugino mass ratios, $a$ and $b$, needed to get low fine tuning. This is shown in Fig.~\ref{fig:7}.
For every shown point we select the smallest fine tuning, i.e.\ the additional parameters are chosen such at each point that
the smallest fine tuning is realised. To connect with our analytical discussion above we not only show the overall fine tuning, but also the fine
tuning with respect to the gaugino mass parameter $m_{1/2}$.
We see that the smallest fine tuning is realised in an elliptic structure in the $a$-$b$ plane. Comparing with Fig.~\ref{MSSManalytic} we see that this structure is the one expected on the basis of the gaugino fixed point. For the fine tuning with respect to $m_{1/2}$ we see that very small values of the fine tuning are possible, $\Delta_{m_{1/2}} \ll 10$. For the overall fine tuning, which is larger and typically dominated by the fine tuning with respect to $\mu$, there is a somewhat broader minimum ring.
For moderate $a$ and before the Higgs mass cut, the optimal value for the magnitude of the Wino to gluino mass at the unification scale is of $|b|=\mathcal{O}(3-4)$. In the simplified analytic calculation of Section~\ref{sec:MSSManalytic} (which neglects $A$ terms) this region corresponds to a gaugino focus point below 10 $\Tev$. Once the Higgs mass cut is imposed the low-fine tuned region requires $a$ and $b$ to lie in the first quadrant, usually with large $a$. 
This is due to the fact that A-terms are also sensitive to $a,b$ and optimal mixing is required for the Higgs mass to be in agreement with experimental results. The numerical results for fine tuning are in agreement with the approximate analytic predictions of Fig.~\ref{MSSManalyticFT1}.

\begin{figure}[!h!]
\centering
\includegraphics[width=0.49\linewidth]{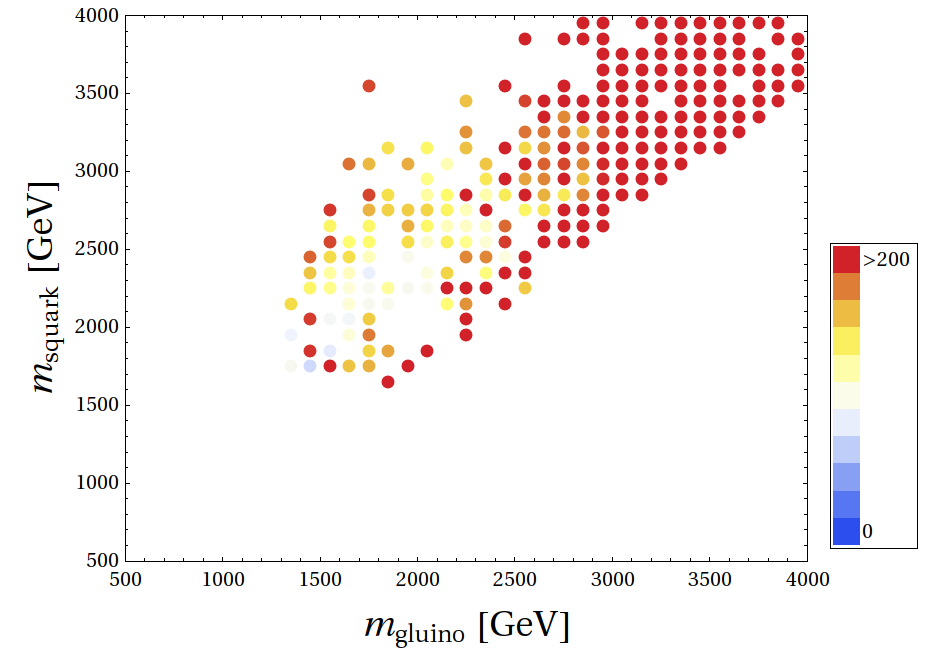}  
\includegraphics[width=0.49\linewidth]{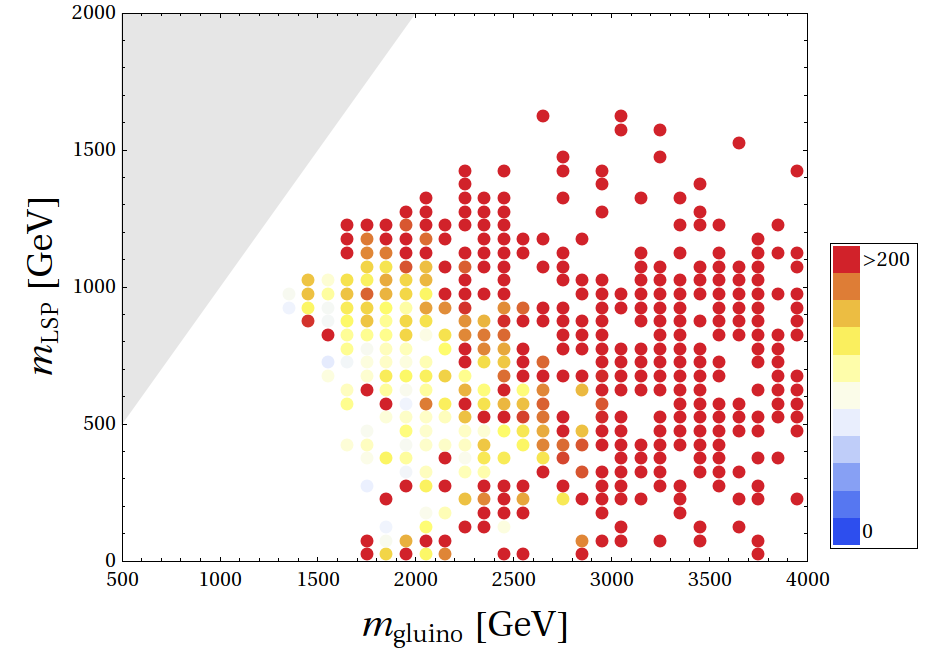}  
\caption{Fine tuning in the gluino-squark and gluino-LSP plane for the MSSM after all cuts.}
\label{fig:mssmpheno}
\end{figure}

In Fig.~\ref{fig:mssmpheno} we show the fine tuning in the gluino-squark and gluino-LSP plane. 
Due to the additional flexibility in the gaugino sector there is the interesting possibility of compressed spectra
where the mass difference between the gluino and the lightest neutralino is not large. 
While in the CMSSM the ratio between $M_1:M_2:M_3$ is about $1:2:6$, for non-universal gaugino masses the Wino and bino can be as heavy as the gluino
where we expect $M_1 \sim M_2\sim M_3$ for $a\sim 6$ and $b\sim 3$. The only other relevant parameter is the $\mu$ term, which sets the mass for the higgsino-like
neutralinos. 
In the MSSM, however, it turns out that in order to achieve the required Higgs mass the gluino mass is much larger than $\mu$, and compressed spectra are typically not realised.
This can also be seen in Fig.~\ref{fig:mssmpheno} as there are no points close to the diagonal. 

\subsection{Dark matter in the MSSM}
 
\begin{figure}[!h!]
\centering
\includegraphics[height=5cm,width=0.42\linewidth]{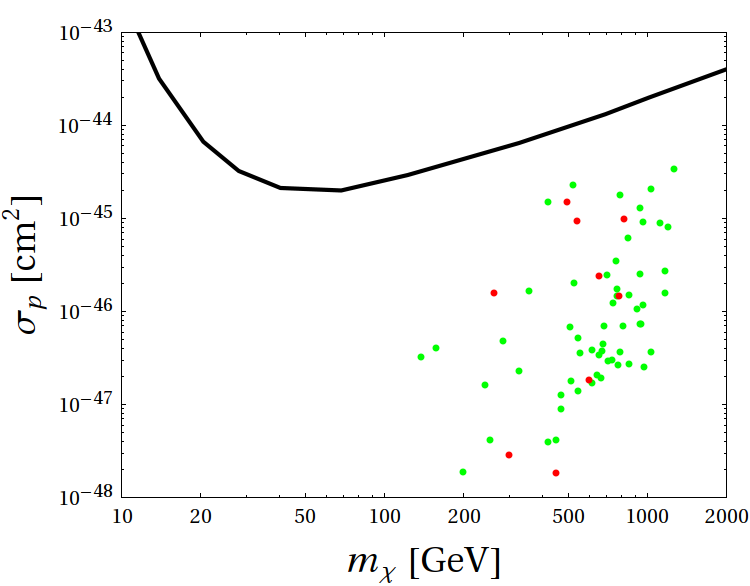}  
\includegraphics[height=5.1cm,width=0.51\linewidth]{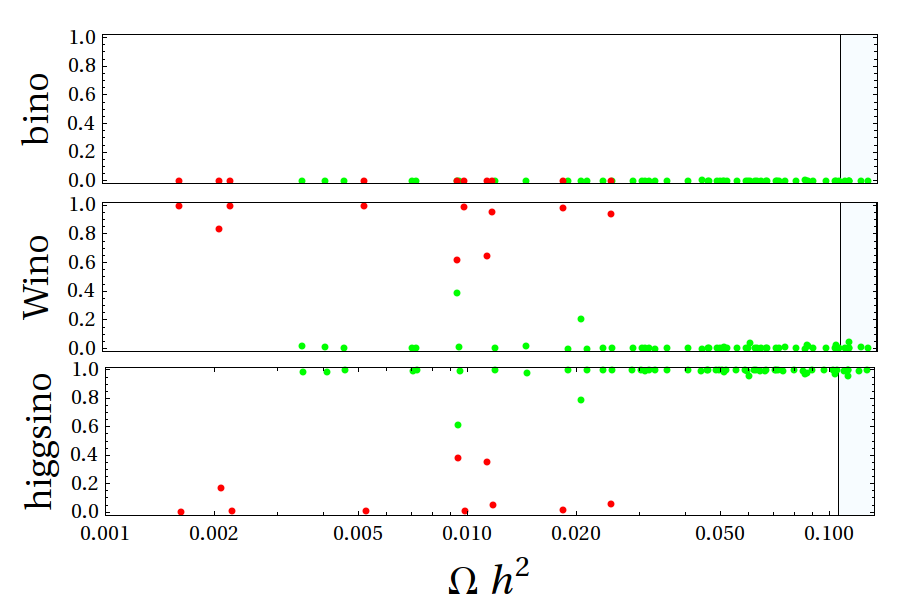}  
\caption{Scaled (i.e.\ multiplied by the ratio of the density of SUSY dark matter to the observed dark matter density, $(\Omega h^2)^\text{th} / 0.1199$) dark matter direct detection cross section as a function of the neutralino mass together with the latest bound from XENON100 \cite{Aprile:2012nq}. Also shown is the dark matter
composition as a function of the relic density. Red points correspond to a mainly Wino-like LSP while green points correspond to a mainly higgsino-like LSP.
For all points, in addition to the SUSY and Higgs cuts, a fine tuning cut, $\Delta < 300$, was imposed.}
\label{fig:8}
\end{figure}

Due to the additional flexibility in the gaugino sector, a large variety of LSP compositions is possible.
To infer something about the typical phenomenology we concentrate on regions corresponding to the smallest fine tuning. As there are no points with particularly small fine tuning
for the MSSM case we look at viable points with a fine tuning below 300. Also, in order to ascribe meaning to the density of points, we
only show points from the scan with a uniform density in the input parameters.
In Fig.~\ref{fig:8} we show the direct detection cross section vs.\ the mass of the lightest neutralino together with the latest bound from XENON100 \cite{Aprile:2012nq} 
as well as the dark matter composition as a function of the relic density.
It can be seen that all of the points are below the XENON100 direct detection limit.
Regarding the composition we see that for points satisfying the relic abundance upper bound the LSP is mainly composed of Wino and higgsino,
with typically only a very small bino component.
For the case of a Wino-like LSP we see that the relic abundance is always below the one required for dark matter.
On the other hand the correct relic abundance is often achieved with a higgsino-like LSP.

\section{Beyond the MSSM - the GNMSSM}
\label{sec:gnmssm}
\subsection{Operator analysis}
Given the present lack of evidence for the MSSM at the LHC and its associated fine tuning problem it is important to ask if  alternative SUSY extensions of the SM have lower fine tuning. A useful way to look for such extensions is to allow for a general modification of the MSSM by adding higher dimension operators that correspond to the effective field theory that results from integrating out additional heavy degrees of freedom and ask if such operators can reduce fine tuning.
There is a unique leading dimension 5 operator with the form\cite{Cassel:2009ps}
\be
L=\frac{1}{M_{*}}\int d^{2}\theta f(X)(H_{u}H_{d})^2 \;,
\label{operator}
\ee
where $X=\theta\theta m_{0}$ and $m_{0}$ is the SUSY breaking scale in the visible sector. 

This gives contributions to the scalar potential of the form
\be
V=(|h_{u}|^{2}+|h_{d}|^{2})(\chi_{1}h_{u}h_{d}+h.c.)+\frac{1}{2}\big(\chi_{2}(h_{u}h_{d})^{2}+h.c.\big)
\label{VX}
\ee
where $\chi_{1}=2f(0)\mu_\text{eff}/M_{*},\;\chi_{2}=-2f'(0)m_{0}/M_{*}$ and $\mu_\text{eff}$ is the effective $\mu$ term including the singlet contribution.

Note that the $\chi_{1}$ term is supersymmetric so there are associated corrections involving Higgsinos that will generate Higgsino mass terms of the same order of magnitude as the correction to the Higgs mass terms (once the Higgs acquire their vevs). However in practice these corrections are going to be of $\mathcal{O}(10 \gev)$, important to get a Higgs mass of $125\gev$ but small compared to the Higgsino mass coming from the $\mu_\text{eff}$ term which will have to be  of $\mathcal{O}(1 \tev)$. For this reason we concentrate on the effect in the scalar sector.

The fine tuning of this model has been analysed in \cite{Cassel:2009ps} where it was shown that the fine tuning is significantly reduced by the first term of Eq.(\ref{VX}) while the second term only gives a modest reduction. The reason for this is because the dominant effect comes from the contribution of Eq.(\ref{VX}) to the Higgs mass after electroweak breaking and, due to the fact that the first term involves an extra power of $h_{u}$, it gives the larger contribution.

The obvious question is what new physics can give rise to the first operator corresponding to this term. The answer is through the integration out of a new heavy gauge singlet or $SU(2)$ triplet superfield coupling to the Higgs sector. Interestingly the operator is {\it not} generated in the NMSSM, the simplest singlet extension of the MSSM, as it requires an explicit mass term for the singlet super field. We refer to the model with the singlet mass term as the generalised NMSSM (the GNMSSM).  

\subsection{The GNMSSM superpotential}
The most general extension of the MSSM by a gauge singlet chiral superfield consistent with the SM gauge symmetry has a superpotential of the form
\begin{eqnarray}
 \mathcal{W} &=& \mathcal{W}_\text{Yukawa}  + \frac{1}{3}\kappa S^3+
(\mu + \lambda S) H_u H_d + \xi S+ \frac{1}{2} \mu_s S^2  \label{gen}\\
\label{eq:wgnmssm}
&\equiv& \mathcal{W}_\text{NMSSM}+
\mu H_u H_d + \xi S+ \frac{1}{2} \mu_s S^2  \label{gen2} 
\end{eqnarray}
where $\mathcal{W}_\text{Yukawa}$ is the MSSM superpotential generating the SM Yukawa couplings and $ \mathcal{W}_\text{NMSSM}$ is the ``normal'' NMSSM with a $\Z{3}$ symmetry. Throughout this article capital letters refer to superfields while small letters refer to the corresponding scalar component.
One of the dimensionful parameters can be eliminated by a shift in the vev $v_s$. 
We use this freedom to set the linear term in $S$ in the superpotential to zero, $\xi=0$. 

The form of Eq.~(\ref{eq:wgnmssm}) seems to make the hierarchy problem much worse as the SM symmetries do not prevent arbitrarily high scales for the dimensionful mass terms. However these terms can be naturally of order the SUSY breaking scale if there is an underlying $\Z{4}^R$ or $\Z{8}^R$ symmetry \cite{Lee:2010gv,Lee:2011dya}.
Before SUSY breaking the superpotential is of the NMSSM form. However after supersymmetry breaking in a hidden sector with gravity mediation soft superpotential terms are generated but with a scale of order the supersymmetry breaking scale in the visible sector characterised by the gravitino mass, $m_{3/2}$. With these the renormalisable terms of the superpotential take the form \cite{Lee:2011dya}
\begin{eqnarray}
 \mathcal{W}_{\Z{4}^{R}}
 & \sim &    \mathcal{W}_\text{NMSSM}+ m_{3/2}^2\, \singlet + m_{3/2}\, \singlet^{2} 
               + m_{3/2}\, \hu\, \hd \;,\\
   \mathcal{W}_{\Z{8}^{R}}
  & \sim & \mathcal{W}_\text{NMSSM}+ m_{3/2}^2\, \singlet 
\label{eq:WNMSSM1}
\end{eqnarray}
where the $\sim$ denotes that the dimensional terms are specified up to $\mathcal{O}(1)$ coefficients.
Clearly the $\Z{4}^R$ case is equivalent to the GNMSSM. After eliminating the linear term in $S$ the $\Z{8}^R$ case gives a constrained version of the GNMSSM with $\mu_{s}/\mu=2\kappa/\lambda$.

Note that the SUSY breaking also breaks the discrete $R$ symmetry but leaves the subgroup $\Z{2}^{R}$, corresponding to the usual matter parity, unbroken. As a result the lightest supersymmetric particle, the LSP, is stable and a candidate for dark matter.

\subsection{Supersymmetry breaking}

The general soft SUSY breaking  terms associated with the Higgs and singlet sectors are
\begin{align}
 V_\text{soft} 
   &=  m_s^2 |s|^2 + m_{h_u}^2 |h_u|^2+ m_{h_d}^2 |h_d|^2 \nonumber \\
   &+ \left(b\mu \, h_u h_d + \lambda A_\lambda s h_u h_d + \frac{1}{3}\kappa A_\kappa s^3 + \frac{1}{2} b_s s^2  + \xi_s s + h.c.\right) \;.
\label{soft}
\end{align}
Note that the shift in the vev $v_{s}$ that is used to eliminate the linear term in the superpotential does not imply that the corresponding soft term $\xi_s$ is zero as well.

These terms and the soft breaking terms associated with the squarks, sleptons and gauginos depend on the details of the supersymmetry breaking sector. Here we will assume 
underlying GUT relations with non-universal gaugino masses which, as above, we write as $M_1=a \cdot m_{1/2}, M_2=b \cdot m_{1/2}$ and $M_3=m_{1/2}$. All other parameters we take to be CMSSM like, i.e.\ a universal scalar mass and all other soft terms proportional to their corresponding superpotential couplings.

The independent supersymmetry breaking parameters are therefore  
$m_0$, $M_1, M_2, M_3$, $A_0$, $B_0$ and $\xi_s$ where $A_0$ and $B_0$ are the constants of proportionality associated with the trilinear and bilinear terms respectively. These parameters are defined at the unification scale, $M_{X}$, and must be evaluated at low scales using the renormalisation group running. Taking into account the supersymmetric parameters as well, the supersymmetry breaking scheme is specified by the following set of parameters
$\mu$, $\mu_s$, $\lambda$, $\kappa$, $m_0$, $m_{1/2}$, $a$, $b$, $A_0$, $B_0$ and $\xi_s$.
Trading $B_0$, $\xi_s$ and $\mu$ for $v$, $\tan\beta$ and $v_s$ via the EWSB conditions,
there are ten parameters defining this model.

\begin{figure}[!h!]
\centering

\includegraphics[width=0.32\linewidth]{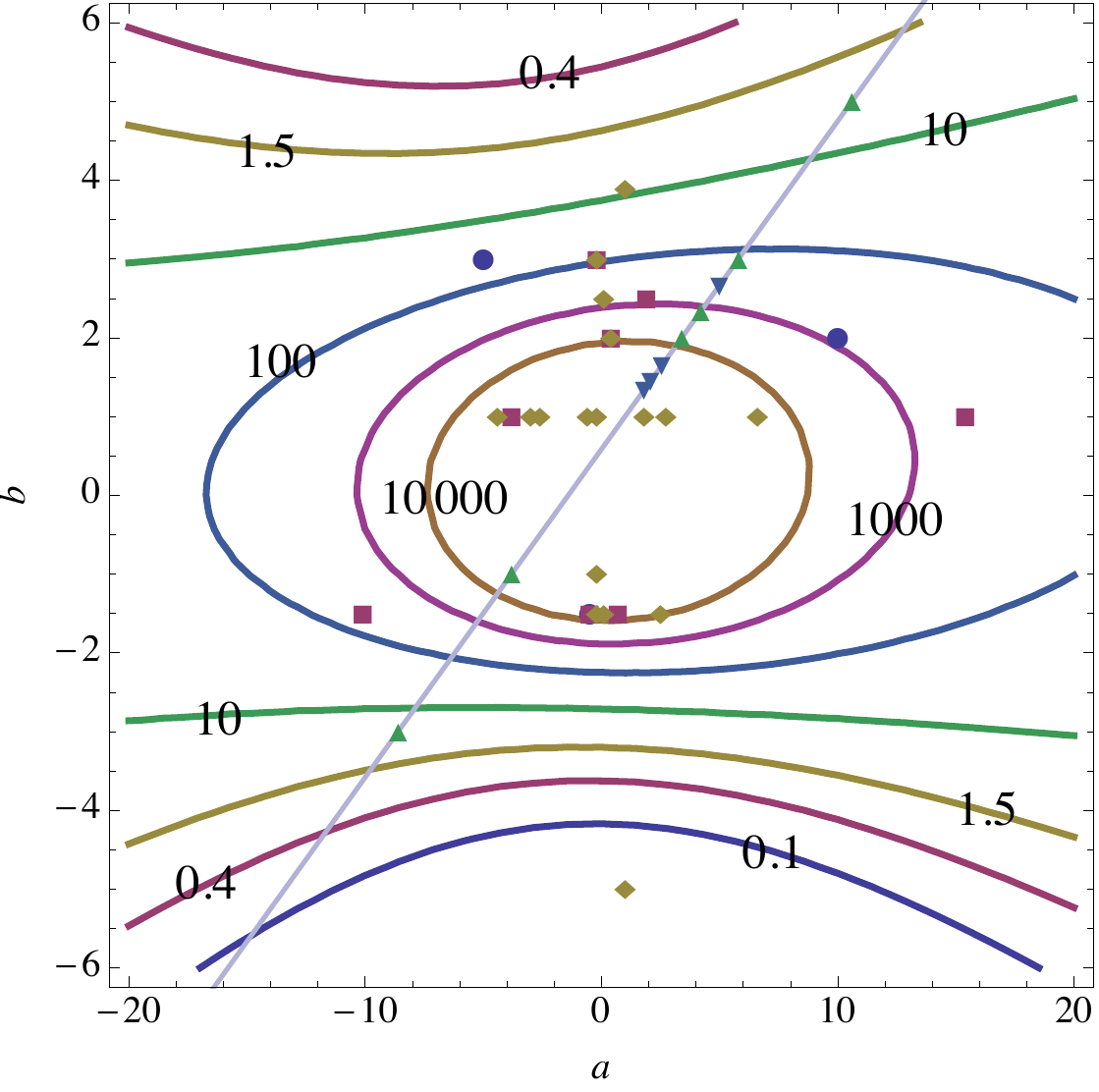}  
\includegraphics[width=0.32\linewidth]{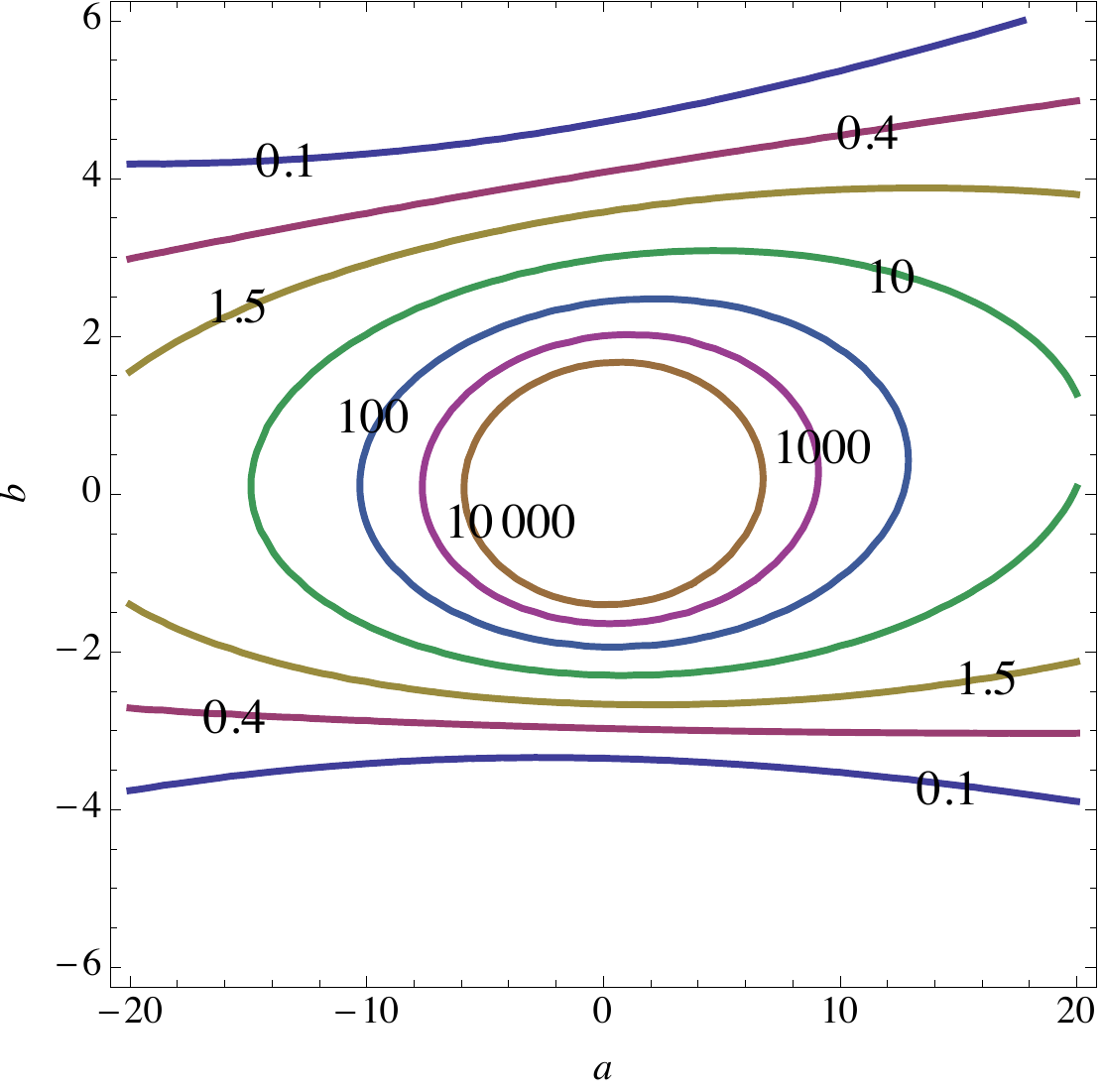}  
\includegraphics[width=0.32\linewidth]{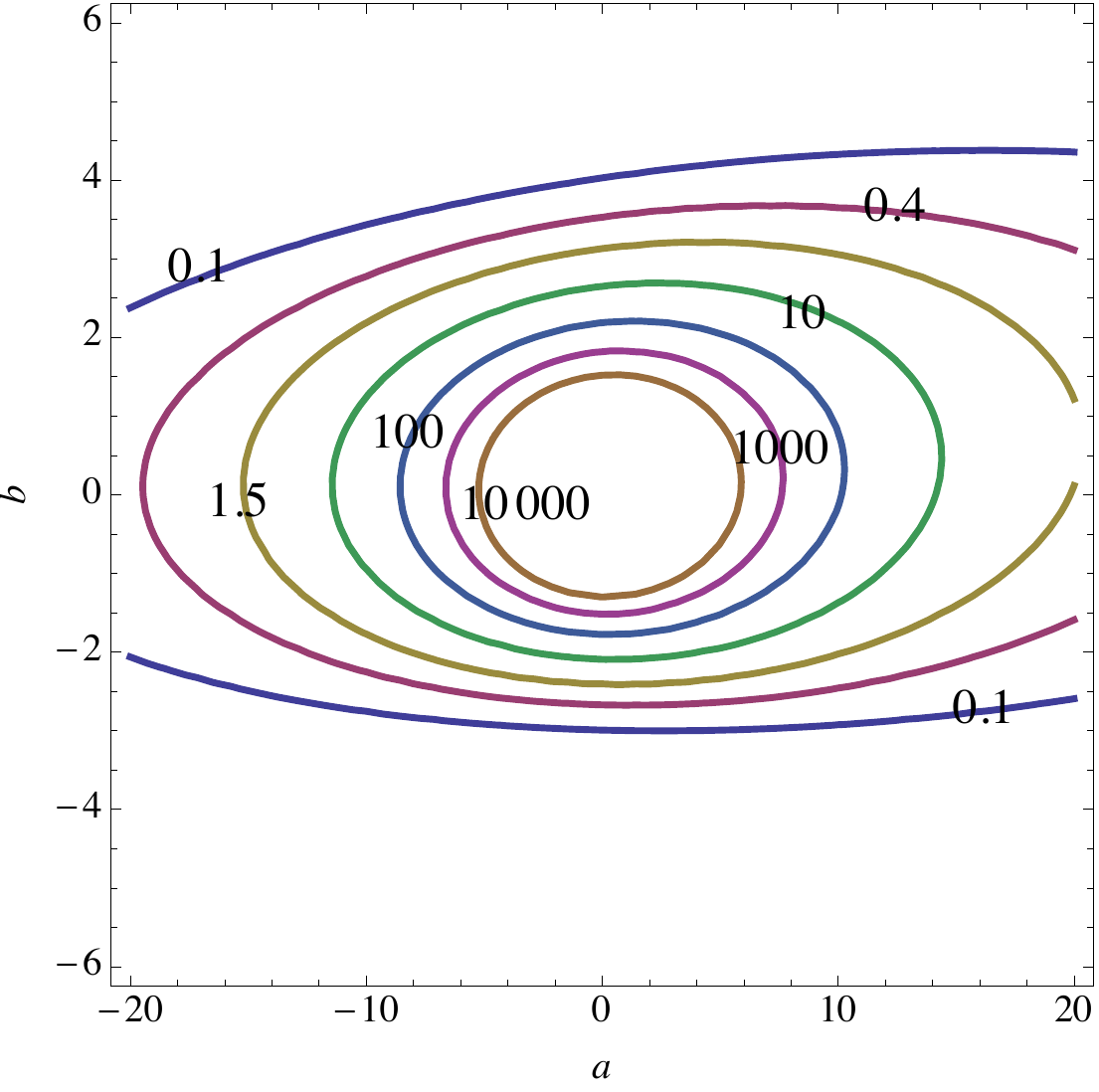}  
\caption{Analytic results for the gaugino focus point scale (in units of $\Tev$) in the GNMSSM for $\lambda=1.5$ (at the GUT scale) and $\tan \beta=2,3,10$ from left to right. The contours do not change much for larger $\tan\beta$. The points and line shown are the same as shown in Fig.~\ref{MSSManalytic}.}
\label{GNMSSManalytic}
\end{figure}

\subsection{The gaugino focus point - analytic discussion}

As before it is instructive to determine the gaugino focus point analytically. Following Section~\ref{sec:MSSManalytic}
we assume that only the top quark Yukawa coupling is non-zero. Due to  the effect of the singlet-Higgs Yukawa coupling, $\lambda$ it is not possible to get an exact solution for the RGE running. Fortunately it is possible to find very good approximate solutions by means of truncated iterations, as suggested in \cite{Mambrini:2001wt}. In the analytic study we also neglect the coupling $\kappa$ as it does not directly affect the Higgs doublets. The results for the gaugino focus point are given in Fig.~\ref{GNMSSManalytic}. The points shown in the first plot correspond to specific models predicting non-universal gaugino masses, as explained in Section~\ref{sec:MSSManalytic}. The general structure is similar to that of the MSSM showing that the focus point is relatively insensitive to $\lambda$.

\subsection{Numerical analysis of fine tuning for the GNMSSM}

\begin{figure}[!h!]
\begin{center}
\includegraphics[width=0.49\linewidth]{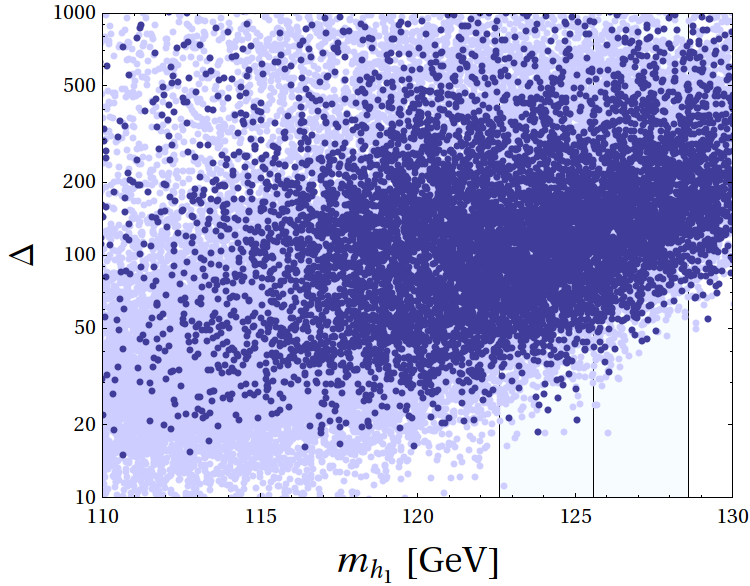}
\includegraphics[width=0.49\linewidth]{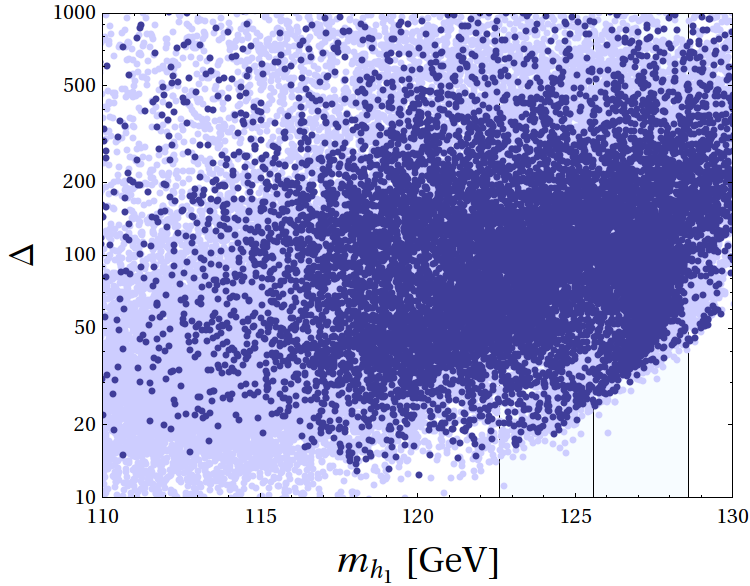}
\caption{Dependence of the fine tuning on the lightest Higgs mass.
The light blue points are before any cuts while the dark blue points take into account
cuts on the SUSY masses and the relic neutralino abundance. The left plot is uniform
in the density of the input parameters, their density reflects the likelihood for finding a viable point.
The right plot shows additional points where we zoomed into regions of small fine tuning. 
The minimal fine tuning after the cuts were imposed is below 20; requiring universal gaugino masses, i.e.\ $a=b=1$, it is about 100.}
\label{CGNMSSMfinetuningmh}
\end{center}
\end{figure}

In the following we will present the results of our full loop-level numerical analysis. Note that this analysis goes beyond the operator analysis of \cite{Horton:2009ed} as we do not require that the singlet mass is large and thus, in general, it cannot be integrated out.
We are particularly interested in regions which allow for a rather large Higgs mass. The largest Higgs masses can be achieved when the additional tree-level contribution
to the Higgs mass is large, corresponding to large $\lambda$, 
(which implies smallish $\kappa$ \cite{Ellwanger:2006rm}) and small $\tan \beta$.
We randomly scan over all the free parameters within this region. In the following the fine tuning is calculated with respect to all independent
high scale input parameters.

Fig.~\ref{CGNMSSMfinetuningmh} shows the fine tuning as a function of the Higgs mass $m_h$.
The minimal fine tuning after the cuts were imposed is below 20; requiring universal gaugino masses, i.e.\ $a=b=1$, it is about 100. Thus we see that there are significant areas of low fine tuning remaining to be explored.
Comparing with Fig.~\ref{GNMSSManalytic} shows that the low fine tuning region corresponds to the gaugino focus point scale close to the electroweak scale for small $\tan\beta$. This pattern is similar to the analytically obtained fine tuning contours in Fig.~\ref{MSSManalyticFT2} corresponding to the low $\tan\beta$ MSSM case without the Higgs mass cut.

\begin{figure}[!h!]
\begin{center}
\includegraphics[width=0.49\linewidth]{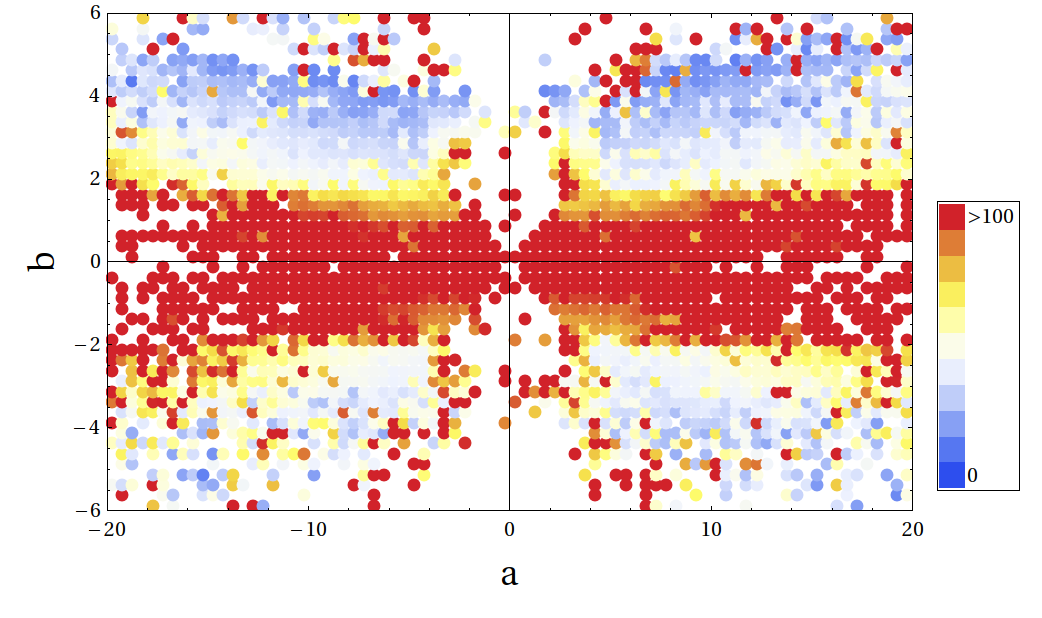}
\includegraphics[width=0.49\linewidth]{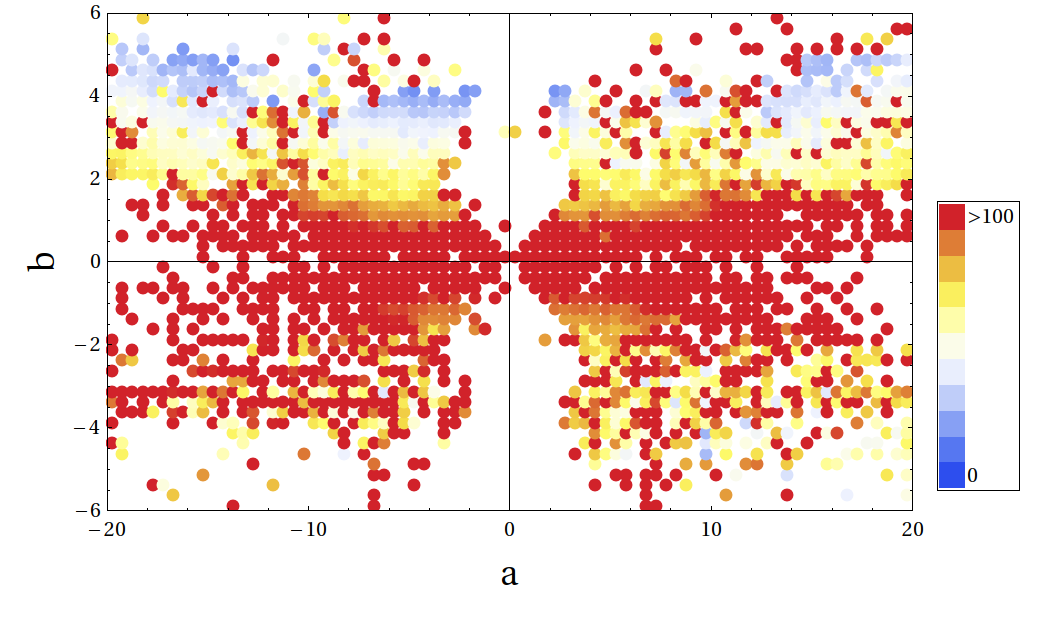}
\includegraphics[width=0.49\linewidth]{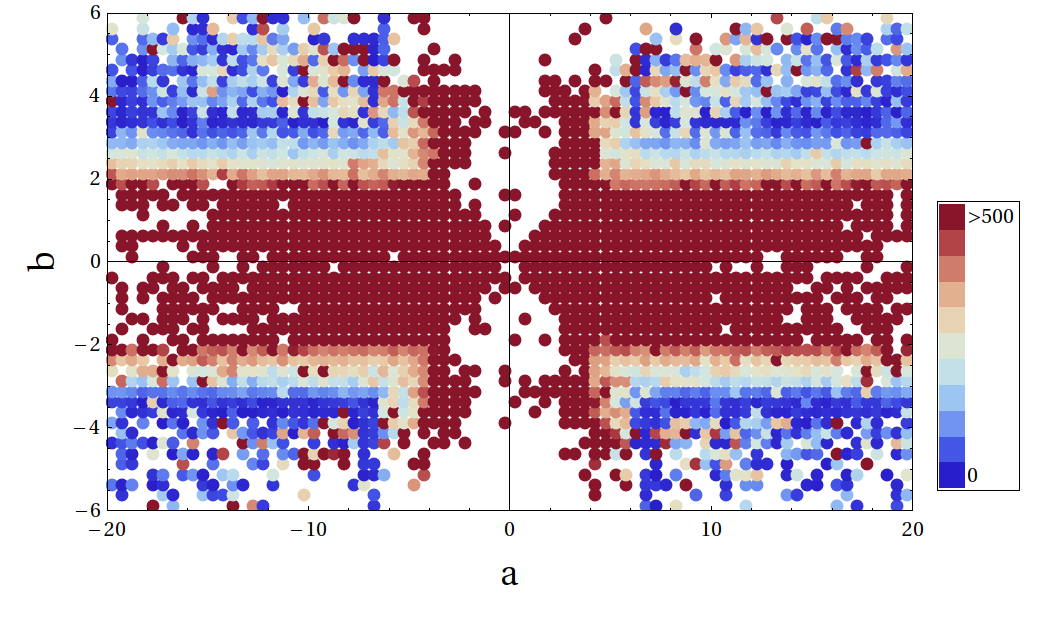}
\includegraphics[width=0.49\linewidth]{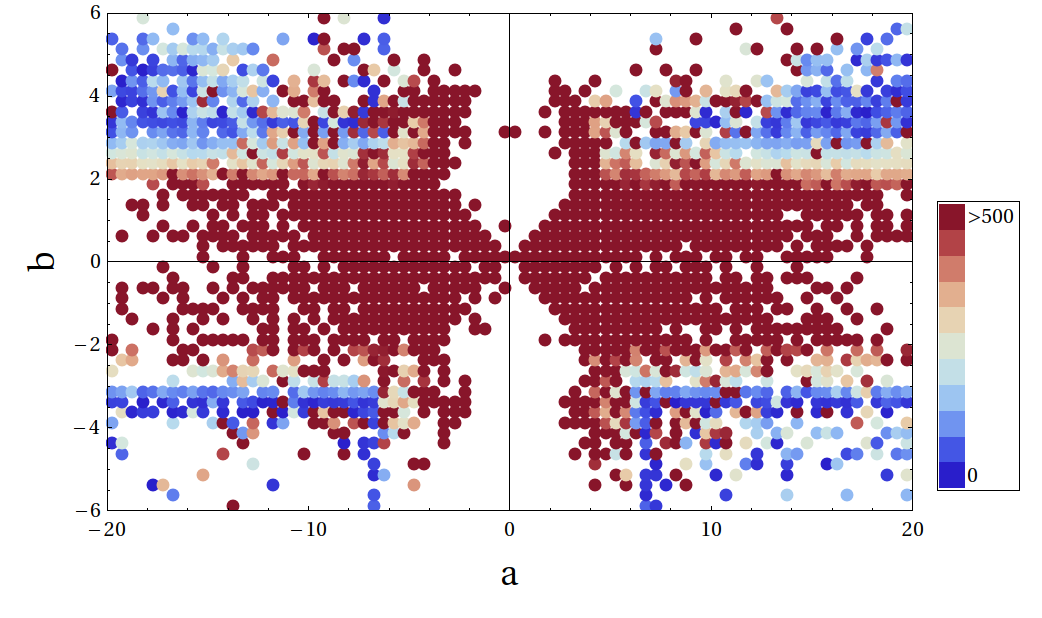}
\caption{Dependence of the fine tuning on the input parameters $a$ and $b$ with the additional parameters chosen such at each point that
the smallest fine tuning is realised.
The upper left plot is after SUSY and DM but before the Higgs mass cut - the upper right plot includes the cut on the Higgs mass.
Note that due to the smaller fine tuning in the GNMSSM the scale is different from the corresponding MSSM figures.
The lower plots show the smallest mass difference in $\Gev$ between the gluino and the neutralino LSP with the same cuts as in the upper figures.
It can be seen that this mass difference can be very small in the regions of low fine tuning, corresponding to very compressed spectra.}
\label{GNMSSMfinetuningab}
\end{center}
\end{figure}

\subsection{GNMSSM phenomenology}
 
\begin{figure}[!h!]
\begin{center}
\includegraphics[width=0.49\linewidth]{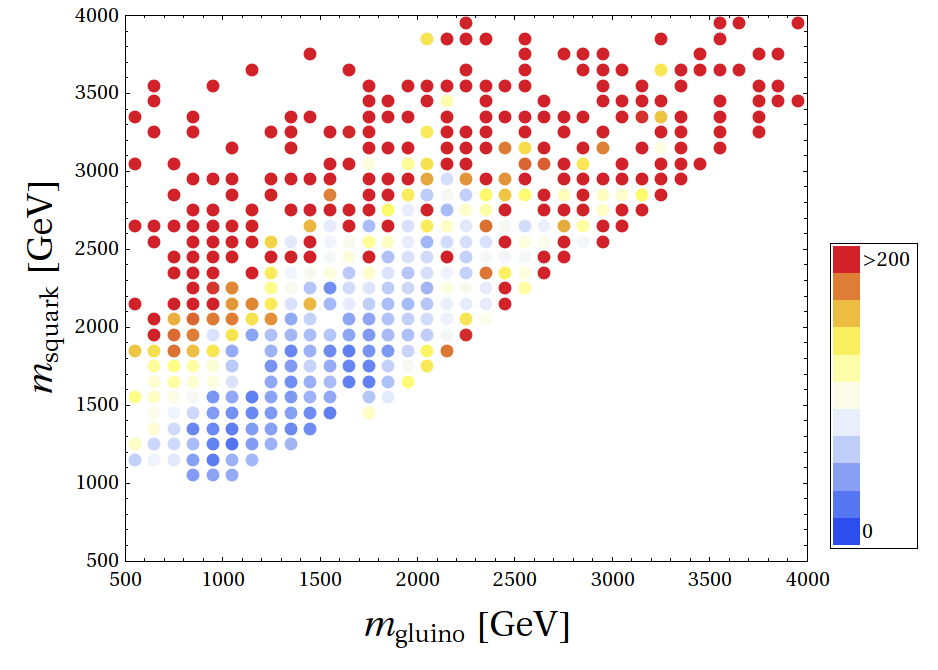}
\includegraphics[width=0.49\linewidth]{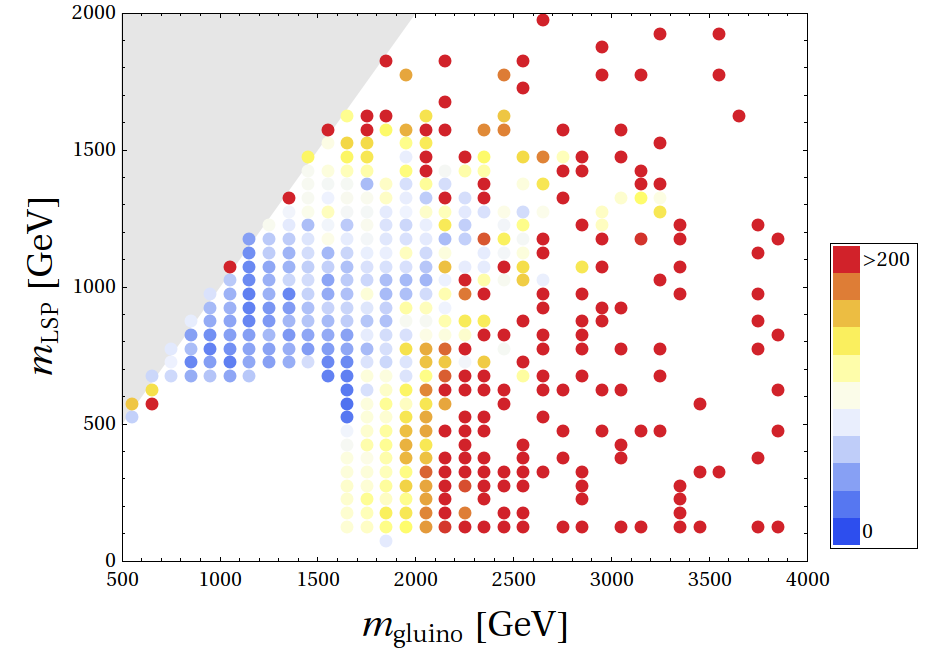}
\caption{Fine tuning in the gluino-squark and gluino-LSP plane for the MSSM after all cuts.}
\label{Pheno}
\end{center}
\end{figure}
To infer something about the typical phenomenology of the low fine tuned regions we show the fine tuning in the gluino-squark and gluino-LSP planes in Fig.~\ref{Pheno}.
It can be seen that points with fine tuning below 100 can have gluino masses beyond $2 \tev$ and squark masses around $3\tev$. 

\subsubsection{Compressed spectra}

As in the MSSM there is the interesting possibility of compressed spectra due to the flexibility in the mass structure of the gaugino sector.
In the MSSM this typically does not happen because often there is  a higgsino-like neutralino considerably lighter than the gluino. However in the GNMSSM  the effective $\mu$ term is often close to $M_3$ so the LSP mass is close to that of the gluino. The resulting compression can be seen in the lower panels of Fig.~\ref{GNMSSMfinetuningab} as well as in the right panel of Fig.~\ref{Pheno}.
For the case of heavy squarks the LHC signal is dominantly gluino pair production and decay but this is significantly reduced for the case of compressed spectra.

\subsubsection{Dark matter}
In the universal scalar mass case considered here the singlet states are always heavy and thus  play no role in the low-energy phenomenology. Their dominant effect is the change to the Higgs mass that reduces the fine tuning, as was found in the CGNMSSM with universal gaugino masses.  However the region of parameter space of the CGNMSSM that solves the little hierarchy problem has essentially been ruled out by a combination of LHC non-observation of SUSY and dark matter abundance. In particular the dark matter abundance has to be reduced below the ``over-closure'' limit and this is dominantly through stau co-annihilation that is only effective for relatively low $m_{0}$ and $m_{1/2}$ and hence requires sparticle masses in the reach of LHC8 \cite{Ross:2012nr}.
For the case of non-universal gaugino masses the situation changes because the LSP can now have a significant non-bino component that allows for its efficient annihilation. As a result the constraint on $m_{0}$ and $m_{1/2}$ is relaxed and there remains a significant part of the low-fine-tuned parameter space to be tested at LHC14. 

\begin{figure}[!h!]
\begin{center}
\includegraphics[height=5cm,width=0.42\linewidth]{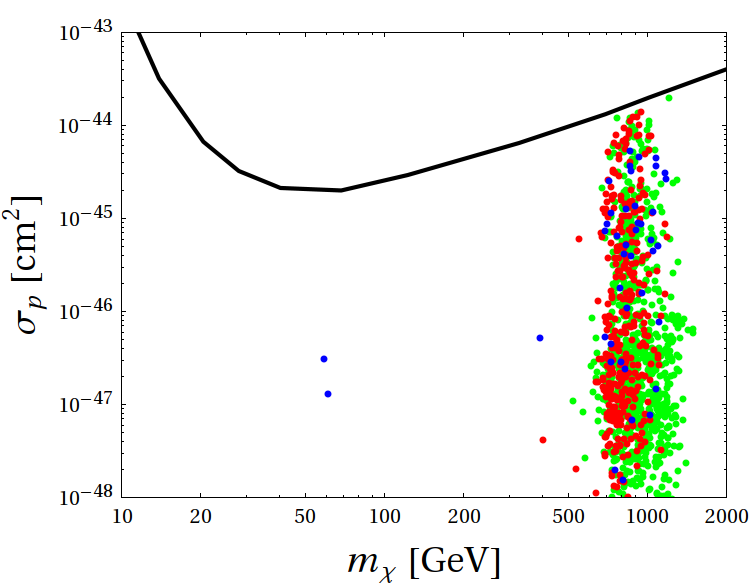}  
\includegraphics[height=5.1cm,width=0.51\linewidth]{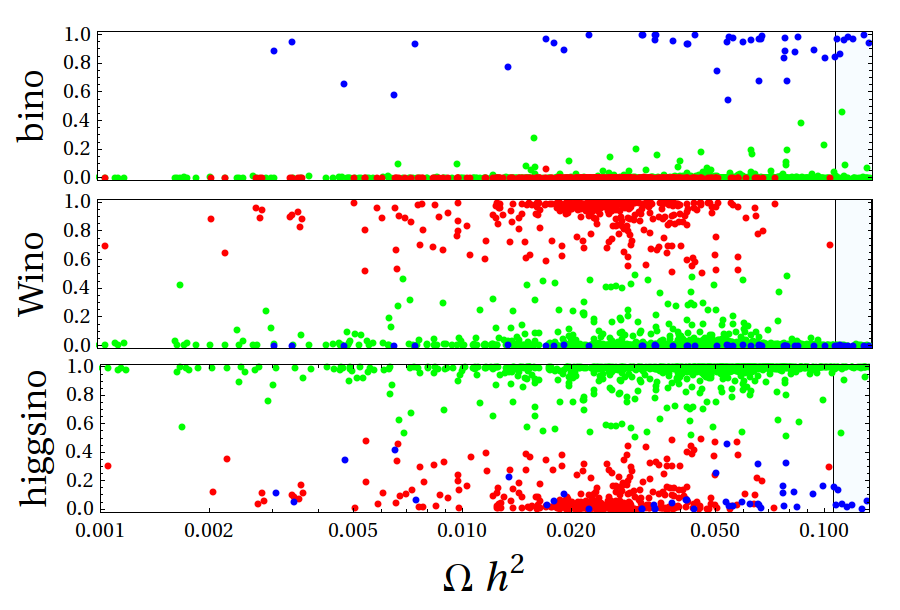}  
\includegraphics[width=0.42\linewidth]{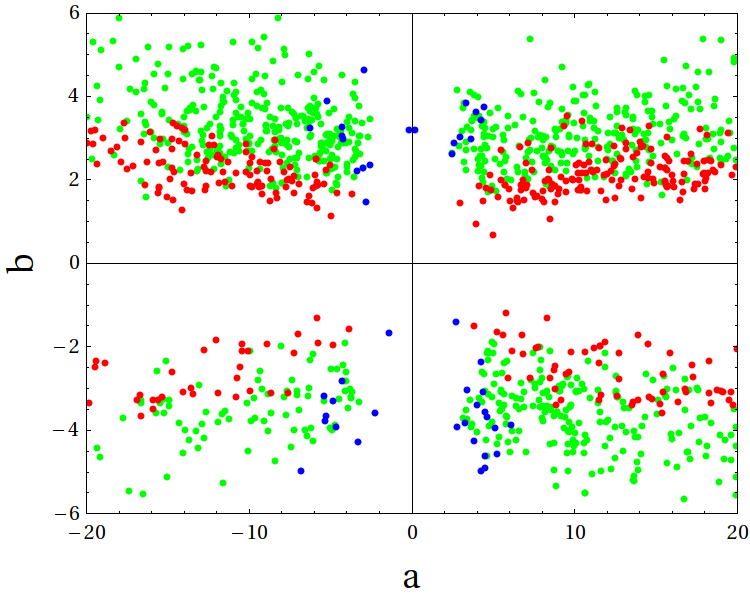}
\caption{(i)The dark matter direct detection cross section as a function of the neutralino mass. It has been scaled (i.e.\ multiplied with $(\Omega h^2)^\text{th} / 0.1199$) to account for cases with underabundant neutralinos. Also shown is the latest bound from XENON100 \cite{Aprile:2012nq}. (ii)The dark matter
composition as a function of the relic density. Mostly bino-like LSPs are shown in blue, mostly Wino-like LSPs are shown in red and mostly higgsino-like LSPs are shown in green. (iii)The distribution of bino-, Wino-, and higgsino like LSPs in the $a$-$b$ plane. For all points, in addition to the SUSY and Higgs cuts, a fine tuning $\Delta < 100$ was imposed.}
\label{GNMSSMDM}
\end{center}
\end{figure}

What about the prospect of direct detection of dark matter?
In Fig.~\ref{GNMSSMDM} we plot, for relatively low-fine-tuned points, the direct detection cross section versus the mass of the lightest neutralino. Also shown is  the latest bound from XENON100 \cite{Aprile:2012nq} as well as the dark matter composition as a function of the relic density. 
In order to ascribe meaning to the density of points, we only show points from the scan with a uniform density in the input parameters. It can be seen that all of the points are below the XENON100 direct detection limit.
Regarding the composition, we see that for the correct relic density or an underabundance the LSP is mainly composed of Wino and higgsino,
with typically only a very small bino component.
As in the MSSM the correct relic abundance seems to be more easily achieved with a higgsino-like LSP.

\subsubsection{Outlook}
Our analysis has concentrated on the more fine-tuned UV complete case with parameters defined at the unification scale and requiring  gauge coupling unification. Even so it is clear from the figures that models with non-universal gaugino masses still have a sizable region of parameter space with reasonably small fine tuning that remains to be tested. While LHC14 will cover most of this region, some points lie beyond  even its reach,
particularly for the case of a compressed spectrum.
In the context of gauge coupling unification an interesting feature of the low fine-tuned regions around $b\sim 3-4$ is that within these regions the precision of gauge coupling
unification is much better than in the case of universal gaugino masses, see e.g.~\cite{Raby:2009sf}.

Similar to the case of the CGNMSSM we find that viable points have a large supersymmetric singlet mass parameter, leading to heavy singlet states. In detail this constraint comes from the need to achieve acceptable electroweak breaking consistent with the universality of scalar masses at the high scale and the observed Higgs mass.  Indeed this is also why universal scalar masses are not possible in the NMSSM. Allowing for non-universal Higgs masses solves this problem for the NMSSM and in the case of the GNMSSM it will allow for light singlet states that can have interesting phenomenological implications within the GNMSSM \cite{SchmidtHoberg:2012yy,SchmidtHoberg:2012ip}. The case of more general boundary conditions will be considered elsewhere.

\section{Summary and Conclusions}
\label{sec:conclusions}
The non-observation, to date, of evidence for the existence of supersymmetric partners of the SM states has cast doubt on the viability of a supersymmetric solution to the hierarchy problem. To quantify this it is useful to use a fine-tuning measure that determines the precision of cancellation needed between uncorrelated parameters. Through the connection of the measure to a likelihood fit to the data one obtains an upper bound significantly less than 100 on the fine-tuning measure, $\Delta$, that is consistent with an acceptable SUSY fit to the data. For the case that one does not require a UV completion of the SUSY model the fine tuning is minimised because of the absence of large renormalisation group logarithmic terms. In this case, even for top squarks with TeV masses, perfectly acceptable values, $\Delta\sim 10$, are obtained for the MSSM \cite{Baer:2012cf}. However if one requires a (GUT) UV completion and the associated gauge coupling unification the fine tuning in the CMSSM becomes unacceptable, $\Delta\ge 350$, largely because it is difficult in the CMSSM to get a Higgs as heavy as $126 \gev$.

In this paper we have determined the fine tuning for a simple extension of the CMSSM,  that allows for non-universal gaugino masses at the unification scale, which preserves gauge coupling unification and is consistent with the LHC bounds on SUSY masses, a $126 \gev$ Higgs and acceptable dark matter abundance. We found that, due to a gaugino ``focus point'', the fine-tuning is considerably reduced with a minimum $\Delta\sim 60$, just acceptable.  We also determined the fine tuning for a further extension of the model, the (C)GNMSSM, that includes a singlet supermultiplet. In this case, due mostly to the additional contribution to the Higgs mass coming from the singlet F-term, the fine tuning is further reduced and can be as low as $\Delta=20$, perfectly acceptable. This means that, even after the LHC8 results, simple SUSY models can still provide a solution to the hierarchy problem that preserves the successful prediction following from gauge coupling unification. 

For the case of the (C)MSSM  the regions of parameter space with $\Delta<100$ have SUSY masses in the reach of LHC14 and a spectrum that gives the usual large missing energy signatures of SUSY. As a result the model will be fully tested by LHC14. However for the (C)GNMSSM there is a significant region of parameter space for which the spectrum is compressed giving signals below the present range of sensitivity of the LHC experiments. Moreover there are regions with quarks and gluinos beyond the reach of LHC14. As a result, while LHC14 will probe most of the remaining low-fine-tuned region of the (C)GNMSSM parameter space with $\Delta<100$, it will not be able to cover the entire region. In both the (C)MSSM and the (C)GNMSSM models higgsino-like dark matter is favoured and the direct dark matter cross section can be significantly below the XENON100 bound.

\section*{Acknowledgements}
We would like to thank Vinzenz Maurer for correspondence and Florian Staub for discussions. The research presented here was partially supported by the EU ITN grant UNILHC 237920 (Unification in the LHC era).
One of us (GGR) would like to thank the Leverhulme
foundation for the award of an emeritus fellowship that also supported this research.

\appendix

\section{Fine tuning in the MSSM}

In the limit of large $\tan\beta$ the fine tuning in the MSSM with respect to initial parameters is given approximately by
\bea
\Delta_{\mu_0} & \approx & -\frac{4\mu^2}{m_Z^2}\frac{1}{1+\delta} \nn \\
\Delta_{m_0} & \approx & -\frac{4m_0^2}{m_Z^2}z_{h_u}^{m_0}\left( v\right) \frac{1}{1+\delta} \nn \\
\Delta_{m_{1/2}} & \approx & -\frac{4m_{1/2}^2}{m_Z^2}z_{h_u}^{m_{1/2}}\left( v\right) \frac{1}{1+\delta}-\frac{4m_{1/2}A_{0}}{m_Z^2}z_{h_u}^{m_{1/2}A_0}\left( v\right) \frac{1}{1+\delta} \nn \\
\Delta_{A_{0}} & \approx & -\frac{4A_{0}^2}{m_Z^2}z_{h_u}^{A_0}\left( v\right) \frac{1}{1+\delta}-\frac{4m_{1/2}A_{0}}{m_Z^2}z_{h_u}^{m_{1/2}A_0}\left( v\right) \frac{1}{1+\delta} 
\eea
where
\begin{equation}
\delta\approx\frac{3h_{t}^4}{g^2\pi^2}\left[ \log\frac{m_{\tilde{t}}}{m_{t}}+\frac{1}{2}\frac{\left( A_{t}-\mu/\tan\beta\right) ^2}{m_{\tilde{t}^2}}\left( 1-\frac{\left( A_{t}-\mu/\tan\beta\right) ^2}{12m_{\tilde{t}^2}}\right) \right] .
\label{delta}
\end{equation}
Taking into account that
\begin{equation}
\frac{1}{2}m_{Z}^2\approx -\left( m_{Hu}^2+\mu^2\right) \frac{1}{1+\delta}
\end{equation}
the fine tuning with respect to $\mu_0$, which is often dominating, can be approximated by
\begin{equation}
\Delta_{\mu_0} \approx 2+\frac{4m_{h_u}^2}{m_Z^2}\frac{1}{1+\delta}
\end{equation}
which shows that reducing $m_{h_u}^2$ by the gaugino focus point helps not only with lowering $\Delta_{m_{1/2}}$, but also with $\Delta_{\mu_0}$. The approximate value of $m_{h_u}^2$ at the EW scale is given by
\bea
m_{h_u}^2\left( v\right) & \approx & -0.108 A_0^2 - 0.030 m_0^2 \nn \\
& + & \left( -2.024 - 0.027 a + 0.006 a^2 - 0.182 b - 
 0.006 a b + 0.214 b^2\right) m_{1/2}^2 \nn \\
 & + & \left( 0.320 + 0.013 a + 0.078 b\right) m_{1/2} A_{0}
\eea
which shows that the fine tuning with respect to $m_0$ will usually play a negligible role, while the FT both with respect to $m_{1/2}$ and $A_{0}$ will vary in the $\left( a,b\right) $ plane.

For low $\tan\beta$, neglecting the dependence of $\tan\beta$ on initial parameters and noticing that $z_{h_d}^A=z_{h_d}^{MA}=0$, the fine tuning can be approximated by
\bea
\Delta_{\mu_0} & \approx & -\frac{4\mu^2}{m_Z^2}\frac{\tan^2\beta-1}{\tan^2\beta-1+\delta\tan^4\beta/\left( 1+\tan^2\beta\right) }\nn \\
\Delta_{m_0} & \approx & -\frac{4m_0^2}{m_Z^2}\frac{\tan^2\beta\ z_{h_u}^{m_0}(v)-z_{h_d}^{m_0}(v)}{\tan^2\beta-1+\delta\tan^4\beta/\left( 1+\tan^2\beta\right) } \nn \\
\Delta_{m_{1/2}} & \approx & -\frac{4}{m_Z^2}\frac{m_{1/2}^2\left( \tan^2\beta\ z_{h_u}^{m_{1/2}}(v)-z_{h_d}^{m_{1/2}}(v)\right) -m_{1/2} A_0\tan^2\beta\ z_{h_u}^{m_{1/2}A_0}(v) }{\tan^2\beta-1+\delta\tan^4\beta/\left( 1+\tan^2\beta\right) } \nn \\
\Delta_{A_{0}} & \approx & -\frac{4}{m_Z^2}\frac{A_{0}^2\tan^2\beta\ z_{h_u}^{A_0}(v)-m_{1/2}A_{0}\tan^2\beta\ z_{h_u}^{m_{1/2}A_0}(v)}{\tan^2\beta-1+\delta\tan^4\beta/\left( 1+\tan^2\beta\right) } .
\eea

\bibliography{NMSSM}
\bibliographystyle{ArXiv}

\end{document}